\documentclass{LMCS}

\usepackage{subfigure,enumerate,hyperref}

\newcommand{\ignore}[1]{}

\newcommand{\code}[1]{{\texttt{#1}}}

\newcommand{\CODE}[1]{
{\code{
\vspace{-.1cm}
\begin{tabbing}
xxx\=xxx\=xxx\=xxx\=xxx\=xxx\=xxx\=xxxx\=xxxx\=xxxx\=xxxx\=xxxx\=xxxx\=xxxx\kill
        #1            \end{tabbing}}
\vspace{-.1cm}
}}

\usepackage{graphicx}

\def\doi{6 (3:6) 2010}
\lmcsheading%
{\doi}
{1--41}
{}
{}
{Oct.~13, 2009}
{Aug.~18, 2010}
{}   

\begin{document}

\title{Automated Verification of Practical Garbage Collectors}

\author[C.~Hawblitzel]
{
         Chris Hawblitzel\rsuper a
}	
\address
{{\lsuper a}%
         Microsoft Research \\
         One Microsoft Way \\
         Redmond, WA  98052 \\
         USA
}
\email{Chris.Hawblitzel@microsoft.com}

\author[E.~Petrank]
{
        Erez Petrank\rsuper b
}	
\address
{{\lsuper b}%
         Dept of Computer Science \\
         Technion \\
         Haifa 32000 \\
         Israel
}
\email{erez@cs.technion.ac.il}
\thanks
{{\lsuper b}
        Part of this work was done while the author
        was on a sabbatical leave at Microsoft Research.
        Supported by THE ISRAEL SCIENCE FOUNDATION (grant No. 845/06).
}

\begin{abstract}
Garbage collectors are notoriously hard to verify, due to their
low-level interaction with the underlying system and the general
difficulty in reasoning about reachability in graphs. Several papers
have presented verified collectors, but either the proofs were
hand-written or the collectors were too simplistic to use on
practical applications. In this work, we present two mechanically
verified garbage collectors, both practical enough to use for
real-world C\# benchmarks. The collectors and their associated
allocators consist of x86 assembly language instructions and
macro instructions, annotated with
preconditions, postconditions, invariants, and assertions.  We used the Boogie
verification generator and the Z3 automated theorem prover to verify
this assembly language code mechanically. We provide measurements
comparing the performance of the verified collector with that of the
standard Bartok collectors on off-the-shelf C\# benchmarks, demonstrating
their competitiveness.
\end{abstract}

\keywords{Garbage Collection, Verification}
\subjclass{D.2.4}

\maketitle

\section{Introduction}\label{sec-intro}

Garbage collectors automatically reclaim dynamically allocated
objects that will never be accessed again by the program. Garbage
collection is widely acknowledged for supporting fast development
of reliable and secure software. It has been incorporated into
modern languages, such as Java and C\#.
Many recent projects have attempted to verify the safety or correctness of garbage collectors.
The goal of this verification is to reduce the trusted
computing base of a system and increase the system's reliability.  This is
particularly important for secure systems based on proof-carrying
code (PCC) \cite{pcc} or typed assembly language (TAL) \cite{tal:popl};
typical large-scale PCC/TAL systems can verify the safety of the
mutator (the program), but not of the run-time system that manages
memory and other resource on the mutator's behalf.  This prevents
untrusted programs from customizing the run-time system. Furthermore,
bugs in the unverified run-time systems could result in security
vulnerabilities that undermine the guarantees promised by PCC and
TAL.

Proving that garbage collectors are safe and correct has been a
challenge.  In this work, we provide the first fully mechanized
correctness proofs of garbage collectors and allocators realistic
enough to run large, off-the-shelf benchmarks.  To make this
verification tractable, we exploit recent advances in automated
theorem proving technology, using the Boogie \cite{Boogie:Architecture}
and Z3 \cite{Z3:Efficient} tools to provide
{\em automated} verification of the correctness properties.  Our key
contribution is the expression of garbage collector specifications
and invariants in a style that allows efficient, automated
verification.

We verify two collectors, both practical enough for use with
real-world C\# benchmarks: a Cheney copying collector
\cite{mins63,chen70}, with a bump allocator; and a mark-sweep
collector \cite{mcca60}, with a local-cache allocator that allows
fast bump-pointer allocation. Both are simple enough to verify, yet
efficient enough to support realistic benchmarks competitively. The
collectors and their associated allocators consist of x86 assembly
language instructions and macro instructions, annotated with preconditions, postconditions,
invariants, and assertions.  These annotations require significant
human effort to write, but once they are written,
the Boogie verification condition generator and the Z3
theorem prover verify the annotated collectors automatically,
with no further human intervention.
The collectors and allocators are
entirely self-contained, relying on no unverified library code, and
the verification relies on only a minimal set of trusted axioms and
definitions describing 32-bit arithmetic, x86 instructions, memory
words, and the interface to the mutator.

We show how to define higher-level abstractions,
particularly abstractions drawn from region-based type systems, in
terms of these trusted axioms and definitions; these higher-level
abstractions provide forms of local reasoning that make automated
verification tractable.  The verification ensures that if an allocation or garbage
collection operation completes, then the physical heap managed by
the allocator and collector faithfully represents the abstract
graph of objects defined by the mutator.  The verification also
ensures that the garbage collector deallocates all objects
unreached during the collection. The verification does not prove
termination; verified collectors or allocators could fail to
terminate because of an infinite loop, or fail to terminate
properly because of a 32-bit integer overflow exception, or an
explicit halt operation. (The allocators and collectors halt if
they run out of memory, or if the mutator relies on a feature not
supported by our collectors, such as multithreading.)

The collectors and allocators include support for objects, arrays,
strings, header words, interior pointers, static data scanning,
stack scanning, object descriptors, stack frame descriptors,
return-address lookup tables, and bit-level data manipulation,
making them realistic enough to support off-the-shelf
single-threaded C\# benchmarks compiled with the Bartok compiler,
using the native Bartok memory layouts and descriptor formats.  To
assess the efficacy of the proposed collectors, we ran the verified
collectors with the Bartok runtime and compared their performance
with the standard Bartok mark-sweep and generational copying
collectors. The verified collectors demonstrated competitive
performance.

The contributions in this paper include:
\begin{enumerate}[(1)]
\item We provide the first mechanically verified garbage collectors that support a real-world object model, including vtables, arrays, object descriptors, stacks, etc.
\item We provide the first mechanically verified garbage collectors that can link to code generated by a real-world, optimizing compiler (Bartok).
\item We demonstrate how to apply {\em automated} verification to garbage collectors, including both copying and mark-sweep garbage collectors. This automation allows scaling the verification to realistic collectors without employing a huge human effort.
\item We propose a simple, efficient, easy-to-verify mark-sweep collector and allocator based on local caches.
\item We provide the first performance measurements of off-the-shelf C\#
benchmarks running on top of verified garbage collectors.
\end{enumerate}

\paragraph{\bf Outline.}
Section \ref{sec-label} discusses previous work on garbage collector
verification.  Section \ref{sec-boogie} describes Boogie and Z3.
Section \ref{sec-miniature-ms} presents a complete example mark-sweep collector and
allocator in the BoogiePL programming language \cite{Boogie:Architecture}, describing the
specification and invariants in detail.  Section \ref{sec-regions} generalizes
Section \ref{sec-miniature-ms}'s ideas to cover copying collectors, borrowing ideas
from region-based type systems, and Section \ref{sec-miniature-copy} presents
a complete example copying collector in detail.
Section \ref{sec-collectors} presents two simple,
yet practical, collectors (and their allocators): a Cheney-queue
copying collector and an iterative mark-sweep collector.  Section
\ref{sec-measurements} shows that the practical collectors perform reasonably well
compared to Bartok's native collectors on a range of off-the-shelf C\#
benchmarks.  Section \ref{sec-conslusion} concludes.

\paragraph{\bf Code availability.}
The garbage collectors were coded in an x86-like subset of the
BoogiePL language; a small tool automatically extracted the x86
instructions, which were assembled and linked with the benchmarks (see
Section \ref{subsec-assembly}).
The complete BoogiePL code for the two practical collectors is
available as part of the public Microsoft Research Singularity
RDK2 source (in ``Source Code'', in the
base/Imported/Bartok/runtime/verified/GCs
directory, which can be browsed without downloading
all of Singularity) at:
\begin{verbatim}
  http://www.codeplex.com/singularity
\end{verbatim}
The Boogie and Z3 tools (April 2008 release), used to verify the two collectors, are available from:
\begin{verbatim}
  http://research.microsoft.com/specsharp/
\end{verbatim}

\section{Background and related work}\label{sec-label}

Hand-written proofs of garbage collector correctness, at least for
abstract models of garbage collectors, go back decades (e.g.,
\cite{dijk76b,doli94,birk04,leva06}). The work of Birkedal {\em et
al} \cite{birk04} is noteworthy for formally proving a Cheney
copying collector correct, rather than a mark-sweep collector, and
emphasizing {\em local reasoning} based on separation logic.
Nevertheless, the local reasoning is used mainly to separate
pieces of the invariant at a coarse
granularity (e.g. separating invariants about forwarded objects
from unforwarded objects); we offer a different perspective on
local reasoning in Section \ref{sec-regions}.

Other work \cite{russ94,gont96,have99,jack98,gogu98,gao07,burd01,coupetgrimal03} has
mechanically proven garbage collector correctness, but only for
mark-sweep collectors, only using abstract models of memory
(for instance, representing the heap as just a mathematical graph
and the root set as just a mathematical set), only using
abstract models of programs rather than programs executable on
real hardware, and (with the exception of Russinoff~\cite{russ94}),
all using interactive theorem provers.
For example, Russinoff~\cite{russ94} and Havelund~\cite{have99}
both mechanically verify the same small (albeit concurrent) mark-sweep algorithm, which consists of just 11 statements.
In addition to the standard annotations required to declare the algorithm's invariants,
both papers also required, as hints to the theorem prover,
many explicit user declarations of lemmas: 55 lemmas
in Havelund~\cite{have99}, and over 100 lemmas in Russinoff~\cite{russ94}.
(Most of these lemmas are necessary because the theorem provers
lack the ability to automatically instantiate variables in definitions and
quantified formulas at useful values.)
By contrast, the small collector presented in Section~\ref{sec-miniature-ms}
requires no user-declared lemmas; a small number of triggering annotations
embedded in the source code and invariants provide the theorem
prover with enough hints for the proof to succeed.

More recently, McCreight {\em et al}
\cite{mccr07} used an interactive theorem prover to verify the
correctness of both mark-sweep and copying collectors written in a
RISC-like assembly language, with a more realistic memory model.
Furthermore, their results are foundational, requiring trust only
in a small Coq proof checker (which is much smaller than Boogie/Z3),
a specification of correctness, and a RISC machine language model.
This required an enormous effort though, relying on over 10000
lines of Coq scripts per collector, and the treatment of the
memory still falls short of what realistic compilers expect: the
collectors assume that every object has exactly two fields, and
there is no stack, no static data area, no object and stack frame
descriptors, and so on.  We adopt McCreight {\em et al}'s
definition of correctness as a starting point for our work.

Several papers \cite{wang01,monn01} use typed regions to implement
type-safe copying garbage collectors; these garbage collectors
copy live data from an old region to a new region, and then
(safely) delete the old region.  Type safety is a weaker property
than correctness, though, and these techniques don't obviously
extend to mark-sweep collection.  We borrow ideas from typed
regions to help us verify our copying collector.

Banerjee {\em et al}~\cite{Banerjee:Region08} also use regions
to aid program verification, providing a flexible set of
region constructors (region union, region intersection, etc.),
and region predicates (region disjointness, region subset, etc.).
Although their programming language may be too high level
to express practical garbage collectors, their region operations
could be useful for GC verification in a lower-level language.
Note that in contrast to typed regions and Banerjee {\em et al}'s
approach, we do not build regions into our logic or language directly;
instead, our garbage collectors construct regions from more primitive first-order logic concepts.

Vechev {\em et al} \cite{vech07} describe how to mechanically fit
prefabricated, high-level garbage collection building blocks
together in a provably correct way, but they do not mechanically
verify the building blocks themselves.  For instance, they
assume that
``The algorithm skeleton is fixed, and the operations performed
by the skeleton are known to be correct. For example, we assume
that basic stop-the-world tracing is implemented correctly
(i.e., the trace procedure marks all the objects that are reachable
from the pending set when it executes without interruptions).''
We expect our work to be complementary, since our techniques
could be used to verify building blocks for garbage collection.

\section{Boogie and Z3}\label{sec-boogie}

BoogiePL~\cite{Boogie:Architecture} is a simple imperative programming language designed to
support automated program verification. It includes pure
(side-effect free) expressions, written in a standard C/C\#/Java
syntax, imperative statements (which may update local variables
and global variables), pure functions, and imperative procedures.
Procedures support preconditions and postconditions, written with
the keywords \verb`requires` and \verb`ensures`, that specify what
must be true upon entry to the procedure and what the procedure
guarantees is true upon exit from the procedure.  Within a
procedure, loop invariants for \verb`while` loops are written with
the \verb`invariant` keyword.  The following example shows a pure
function \verb`Pos`, which returns true if its argument is
positive, and a procedure \verb`IncreaseX` that adds a positive number
\verb`y` to a global variable \verb`x`:

%\newpage
\begin{verbatim}
  function{:expand true} Pos(i:int)returns(bool){i>0}
  var x:int;
  procedure IncreaseX(y:int)
    requires Pos(y);
    modifies x;
    ensures  x > old(x);
  {
    x := x + y;
  }
\end{verbatim}
In this example, the expression \verb`old(x)` refers to the value of \verb`x` at the beginning of the procedure's execution, so that the postcondition ``\verb`ensures  x > old(x);`'' says that \verb`x` will have a larger value upon exit from the procedure than upon entry to the procedure.  A procedure must disclose all the global variables it modifies (just \verb`x` in this example); this allows callers of the procedure to know which variables remain unmodified by the procedure.  The \verb`expand true` annotation turns a function definition into a macro that is expanded to its definition whenever it is used, so that ``\verb`requires Pos(y);`'' is just an abbreviation for ``\verb`requires y > 0;`''.  (Recursive or mutually recursive macro definitions are disallowed.)

Our programs occasionally use the statement ``\verb`assert P;`'', which asks the verifier to prove \verb`P`, which is then used as a lemma for subsequent proving.  (We do not use the statement ``\verb`assume P;`'', which introduces a new lemma \verb`P` {\em without} proof, since this would make our verification unsound.)

The Boogie tool generates verification conditions from the
BoogiePL code.  These verification conditions are logical formulas
that, if valid, guarantee that each procedure call satisfies the
procedure's precondition, each procedure guarantees its
postcondition, and each loop invariant holds on entry to the loop
and is maintained by each loop iteration.
For example, the verification condition for the \verb`IncreaseX` example above might be:
\begin{verbatim}
  Pos(y) ==> x + y > x
\end{verbatim}
(Here, \verb`==>` is Boogie's syntax for logical implication.)

Boogie passes these
verification conditions to an automated theorem prover, which
attempts to prove the validity of the verification conditions.  We
use the Z3 theorem prover~\cite{Z3:Efficient}, which is efficient, scales to
large formulas, and reasons about many useful first-order logic
theories, including integers, bit vectors, arrays, and
uninterpreted functions.

Both Boogie and Z3 are part of the trusted computing base
for the verified garbage collectors.  In other words, a bug in
Boogie or Z3 could incorrectly lead to a buggy garbage collector being declared ``verified''.
Currently, our trust in Boogie and Z3 rests on the large
amount of testing that they have endured
(including testing at public competitions~\cite{SMTCOMP08}).
In the future, we may also be able to leverage Z3's recent proof
generation feature~\cite{Z3:Proofs},
which generates proofs checkable with a smaller trusted computing base,
although the time and memory overheads of proof generation may be prohibitive.

BoogiePL's data types are more purely mathematical than the data types
in conventional programming languages.  The type \verb`int` represents
mathematical integers, ranging from negative infinity to positive infinity,
while \verb`bv32` represents 32-bit values.  The theorem prover
support for \verb`int` is more mature and efficient than for \verb`bv32`,
so we used \verb`int` wherever possible (Section \ref{sec-collectors} describes how
we reconciled this approach with the x86's native 32-bit words).

BoogiePL also supports array types \verb`[int]t` for any element type \verb`t`,
defining arrays as simple mappings from mathematical integers to elements.
The BoogiePL ``select'' expression \verb`a[i]` retrieves element \verb`i` from array \verb`a`,
where \verb`i` can be any integer.  The BoogiePL ``update'' expression
\verb`a[i := v]` generates a new array, equal to \verb`a` except at element \verb`i`,
where the new array contains the value \verb`v`, so that \verb`(a[i := v])[i] == v` is true for
any \verb`a`, \verb`i`, and \verb`v`.  For convenience, the statement ``\verb`a[i] := v;`'' is
an abbreviation for ``\verb`a := (a[i := v]);`''.  Arrays can also be multidimensional: an
array \verb`a` of type \verb`[int,int]t` supports a select expression \verb`a[i1,i2]`
and an update expression \verb`a[i1,i2 := v]`.  Note that BoogiePL arrays lack many properties
of say, Java arrays.  For example, BoogiePL arrays are not references, so there's no issue
of aliasing: the statement ``\verb`a := b;`'' assigns a copy of array \verb`b` to variable \verb`a`.

Due to formatting constraints, the BoogiePL code shown in this paper
omits most type annotations.  We abbreviate \verb`a<=b && b<c` as \verb`a<=b<c`, and \verb`function{:expand true}` as \verb`fun`.
The notation ``$\forall^{\tt T}$'' is an abbreviation
for the universal quantifier ``$\forall$'' with a particular {\em trigger} ``\verb`T`'',
used as a hint to Z3, as
described further in Section \ref{subsec-trigger}.  For now,
the reader may ignore the ``\verb`T`''.
Finally, the code uses a convention that variables prefixed with
a dollar sign (e.g. ``\verb`$x`'') are ``ghost'' variables,
erased before run-time, as described further in Section \ref{sec-conabs}.

\section{A miniature mark-sweep collector in BoogiePL}\label{sec-miniature-ms}

This section presents a miniature allocator and mark-sweep collector written in the BoogiePL programming language,
introducing some of the invariants used by the more realistic collectors in subsequent sections.
The allocator and collector are implemented as a single BoogiePL file, shown in its
entirety in Figures \ref{fig:exampleDefs}-\ref{fig:exampleAlloc}.
As in previous verified collectors, a large fraction of the code consists of
preconditions, postconditions, loop invariants, and auxillary definitions.
These require human effort to write, but once written, verification is fast and automated.
When run on this example garbage collector,
Boogie verifies all 7 procedures in the collector in less than 2 seconds; since
Boogie and Z3 process BoogiePL files entirely automatically, no human assistance or proof scripts are required:

\begin{verbatim}
  \Spec#\bin\Boogie.exe mini-ms.bpl

  Boogie program verifier version 0.90,
  Copyright (c) 2003-2008, Microsoft.

  Boogie program verifier finished
  with 7 verified, 0 errors
\end{verbatim}
The miniature collector assumes that every object has exactly two fields, numbered \verb`0` and \verb`1`,
and each field holds a non-null pointer to some object.  The collector manages memory addresses
in the range \verb`memLo`...\verb`memHi - 1`, where \verb`memLo` and \verb`memHi` are constants
such that \verb`0 < memLo <= memHi`, but whose values are otherwise unspecified (see Figure \ref{fig:exampleDefs}).
Memory is object addressed, rather than byte addressed or word addressed, so that each
memory location in the range \verb`memLo`...\verb`memHi - 1` contains either an entire object,
or free space big enough to allocate an object in.
The variable \verb`Mem`, of type \verb`[int,int]int`,
represents all of memory; for each address \verb`i` in the range
\verb`memLo`...\verb`memHi - 1` and field \verb`field` in the range 0...1,
the value \verb`Mem[i,field]` holds the contents of the field \verb`field` in the
object at address \verb`i`.
For conciseness, Figure~\ref{fig:exampleDefs} defines \verb`memAddr(i)` to mean
\verb`memLo <= i < memHi`.

The allocator and collector use a variable \verb`Color` to represent the state of
memory at each address.  If \verb`Color[i]` is \verb`0`, the memory at address \verb`i`
is free.  Otherwise, the memory is occupied by an object and is either colored
white (\verb`Color[i] == 1`), gray (\verb`Color[i] == 2`), or black (\verb`Color[i] == 3`).

\subsection{Concrete and abstract states}\label{sec-conabs}

\begin{figure}
\includegraphics[width=0.75\linewidth]{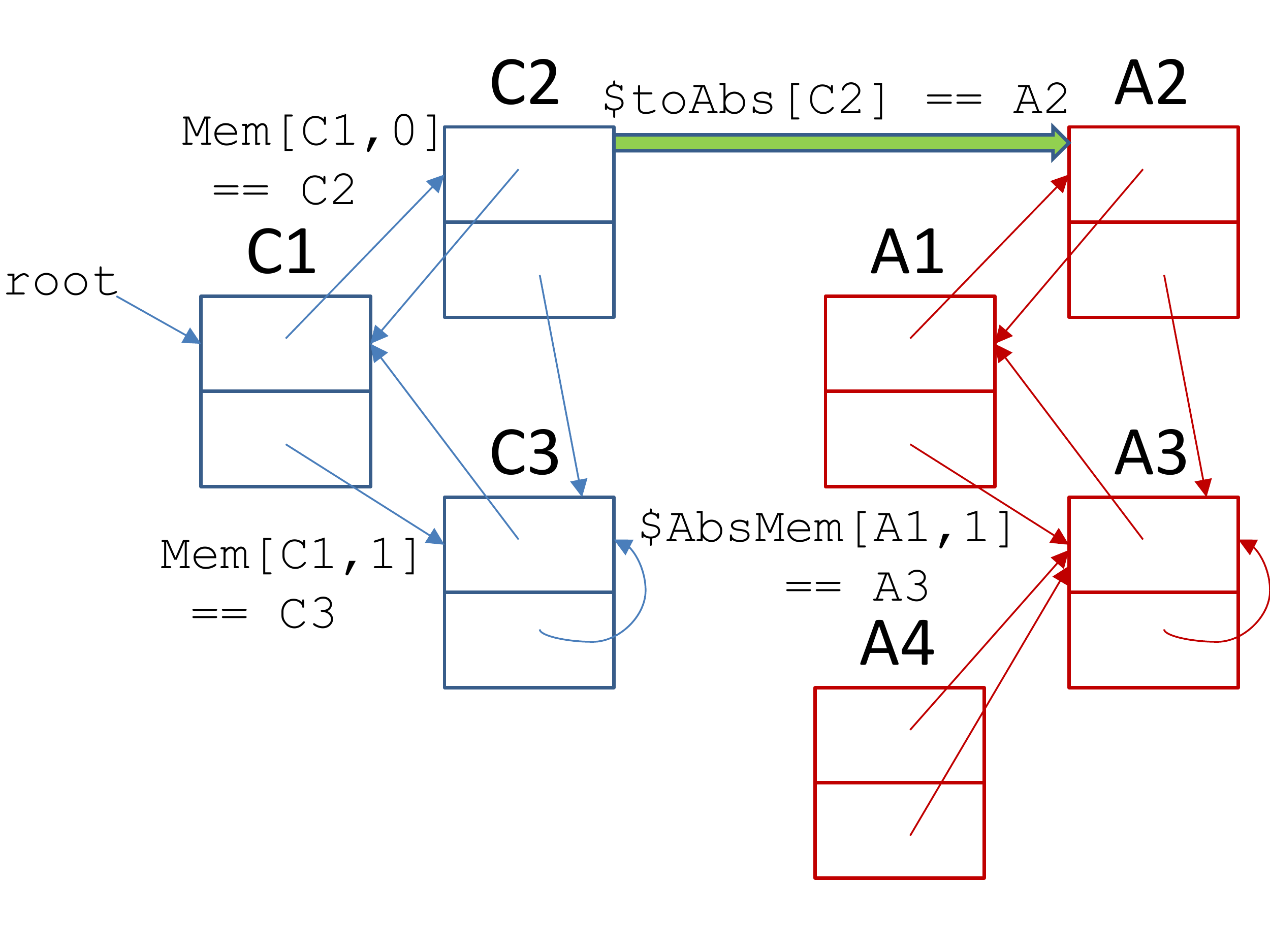}
\caption{Concrete and abstract graphs}\label{fig:conabs}
\end{figure}

To verify a garbage collector, we must specify what it means for a
collector to be correct.  For the mark-sweep collector, the most
obvious criterion is that it frees all objects unreachable from
the root and leaves all reachable objects unmodified.  However,
this definition of correctness is specific to one particular class
of collectors; it doesn't account for collectors that move
objects, and doesn't account for mutator-collector interaction,
such as write barriers and read barriers.  We'd like one
definition of correctness that encompasses many classes of
collectors, so we follow a more general approach advocated by
McCreight {\em et al} \cite{mccr07}.  In this approach, the mutator defines an
abstract state, consisting of an abstract graph of abstract nodes.
A memory manager is responsible for representing the abstract
state in memory.  The memory manager exposes procedures to
initialize memory, allocate memory, read memory, and write memory
(see \verb`Initialize`, \verb`Alloc`, \verb`ReadField`, and \verb`WriteField` in Figures \ref{fig:exampleInit},
\ref{fig:exampleAlloc}).
These four procedures define the boundary between the memory manager and the mutator.
The preconditions and postconditions for these four procedures
express the specification of memory manager's correctness,
where correctness means that each of these procedures faithfully
represents the abstract state.

To make this notion of correctness precise, the variable \verb`$AbsMem` of type \verb`[int,int]int` defines the abstract state as a mapping from abstract nodes and fields to abstract values.  In the miniature memory model presented so far, each field contains a pointer to a node, so the abstract values stored in the abstract graph are always abstract nodes.  (Section \ref{sec-collectors} extends the set of abstract values with other values, such as primitive integers and null.)  For example, Figure \ref{fig:conabs} shows an abstract graph consisting of 4 nodes, \verb`A1`, \verb`A2`, \verb`A3`, and \verb`A4`, each having two fields numbered \verb`0` (on top) and \verb`1` (on the bottom).  In this example, \verb`A1`'s bottom field points to \verb`A3`, so \verb`$AbsMem[A1,1] == A3`.  Integers represent abstract nodes, but these integers can be any mathematical integers, and need not be related to the addresses used by the computer's actual memory.  In fact, the variable \verb`$AbsMem` is not represented at run-time at all; it is used solely for verification.  We call such variables ``ghost variables'' (also known as ``auxillary variables''), and we use a naming convention that prefixes each ghost variable with a dollar sign.

The function \verb`MutatorInv(...)` defines the invariant that holds on the memory manager's data while the mutator is running.  Initialize establishes \verb`MutatorInv`, while \verb`Alloc`, \verb`ReadField`, and \verb`WriteField` require \verb`MutatorInv` as a precondition and guarantee \verb`MutatorInv` as a postcondition.  Each collector defines \verb`MutatorInv(var1...varn)` as it wishes.  The mutator is not allowed to modify any of the variables \verb`var1`...\verb`varn` directly, but instead must use \verb`ReadField`, \verb`WriteField`, and \verb`Alloc` to affect these variables.  Since \verb`MutatorInv` varies across collectors, a mutator that wants to work with all collectors should treat \verb`MutatorInv` as abstract.  In this framework, the specifications for \verb`Initialize`, \verb`Alloc`, \verb`ReadField`, and \verb`WriteField` are exactly the same across all collectors, except for the differing definitions of \verb`MutatorInv`.

The function \verb`$toAbs:[int]int` maps each concrete memory address in the range
\verb`memLo`...\verb`memHi - 1` to an abstract node, or to \verb`NO_ABS`.
The memory management procedures ensure that \verb`$toAbs`
is well formed (\verb`WellFormed($toAbs)`),
which says that any two distinct concrete addresses \verb`i1` and \verb`i2`
map to distinct abstract nodes,
unless they map to \verb`NO_ABS`.
(Note: we use a concrete-to-abstract mapping, rather than an abstract-to-concrete mapping,
because our invariants quantify over concrete addresses, not abstract addresses,
and these quantified concrete addresses make convenient arguments to \verb`$toAbs`.)
In Figure \ref{fig:conabs}, \verb`$toAbs` maps addresses \verb`C1`, \verb`C2`, and \verb`C3` to abstract nodes \verb`A1`, \verb`A2`, and \verb`A3`, respectively, while all other concrete addresses map to \verb`NO_ABS`.   The function \verb`Pointer($toAbs,ptr,$abs)` says that \verb`$toAbs` maps the concrete address \verb`ptr` to the abstract node \verb`$abs`.

Suppose the mutator calls \verb`ReadField(C1,0)`,
which will return the contents of field \verb`0` of the object at address \verb`C1`.
The precondition \verb`Pointer($toAbs,ptr,$toAbs[ptr])`
requires \verb`C1` to be a valid pointer,
mapped to some abstract node (\verb`A1` in this example).
In the miniature memory model presented so far,
all fields hold pointers, so the return value will also be a pointer;
the postcondition for \verb`ReadField` ensures that
the returned value is the pointer corresponding to the abstract node
\verb`$AbsMem[$toAbs[ptr],field]` = \verb`$AbsMem[A1,0]` = \verb`A2`.
Since only one pointer, \verb`C2`, maps to \verb`A2`,
the postcondition forces \verb`ReadField(C1,0)` to return exactly the address \verb`C2`.
(The well-formedness condition, \verb`WellFormed($toAbs)` ensures
that no node other than \verb`C2` maps to \verb`A2`.)
Once the mutator obtains the pointer \verb`C2` from \verb`ReadField(C1,0)`,
it may call, say, \verb`ReadField(C2,1)` to obtain the pointer \verb`C3`.
In this way, the specification of \verb`ReadField` allows the mutator to
traverse the reachable portion of memory,
even though the specification never mentions reachability directly.
The specification does not obligate the memory manager to retain unreachable objects.
Since \verb`A1`, \verb`A2`, and \verb`A3` do not point to \verb`A4`,
the memory manager need not devote any physical memory for representing \verb`A4`.
In Figure \ref{fig:conabs}, there is no concrete address that maps to \verb`A4`.

Note that Figure~\ref{fig:exampleInit}'s implementation of
\verb`ReadField` always returns a value \verb`val` that equals \verb`Mem[ptr,field]`.
Therefore, we could write an alternate version of \verb`ReadField`
that didn't bother to return a value \verb`val`, and instead
wrote ``\verb`Mem[ptr,field]`'' in its postconditions in place of \verb`val`
(e.g. \verb`ensures Pointer($toAbs,Mem[ptr,field],...)`).
In this case, the mutator could perform the load from \verb`Mem[ptr,field]` itself,
relying on \verb`ReadField`'s postcondition to ensure that \verb`Mem[ptr,field]`
corresponds to the proper abstract node.
Similarly, we could write an alternate version of \verb`WriteField`
that omitted the statement \verb`Mem[ptr,field] := val`,
and instead wrote \verb`Mem[ptr,field := val]` in place of \verb`Mem` in its postconditions.
In this case the mutator could store \verb`val` to \verb`Mem[ptr,field]` itself,
relying on \verb`WriteField`'s postconditions to ensure that \verb`MutatorInv`
holds after the store.
In fact, our practical garbage collectors use these alternate versions of \verb`ReadField`
and \verb`WriteField`, so that the practical mutators can inline the loads and stores
(this avoids the run-time overhead of making calls to \verb`ReadField`
and \verb`WriteField` to perform the loads and stores.)

The mutator allocates new abstract nodes by calling \verb`Alloc` (Figure~\ref{fig:exampleAlloc}),
passing in a fresh abstract node \verb`$abs` whose fields initially point to itself.
(A ``fresh'' abstract node is an abstract node that does not yet appear in the range of \verb`$toAbs`.)
Unlike \verb`ReadField` and \verb`WriteField`, \verb`Alloc` modifies \verb`$toAbs`,
which potentially invalidates any pointers that the mutator possesses.
(The mutator can't use an invalid pointer that refers to an old version of \verb`$toAbs`,
because \verb`Pointer($toAbs,...)` for an old \verb`$toAbs`
won't satisfy the preconditions for \verb`ReadField` and \verb`WriteField`,
which are in terms of the current \verb`$toAbs`.)
Therefore, the mutator may pass in a root pointer,
and the \verb`Alloc` procedure returns a new root pointer
that points to the same abstract node as the old pointer.
We could also allow \verb`ReadField` and \verb`WriteField` to modify \verb`$toAbs`,
in which case these procedures would also require a root (or roots) to be passed in.
In practice, though, this would be an onerous burden on the mutator.

\subsubsection{Verifying collection effectiveness}

The specification described so far hides the garbage collection process behind the \verb`Initialize`, \verb`ReadField`, \verb`WriteField`, and \verb`Alloc` interfaces.  We also verify one internal property of the garbage collector, invisible to the mutator: after a collection, only abstract nodes that the collector reached have physical memory dedicated to them; unreached abstract nodes are not represented in memory.  It's easy to define an axiom for reachability for any particular abstract graph: for any node \verb`A`, if \verb`A` is reachable, then \verb`A`'s children are also reachable.  It's difficult, though, to track reachability as the edges in a graph evolve.  For the two collectors presented here, the \verb`$AbsMem` graph remains unmodified throughout collection, but in general, this is not true: incremental collectors interleave short spans of garbage collection with short spans of mutator activity, and the mutator activity modifies \verb`$AbsMem`.  Therefore, we adopt a looser criterion: rather than checking that all remaining allocated nodes at the end of a collection are {\em reachable} from the root, we merely check that all remaining allocated nodes were {\em reached} from the root at some time since the start of the collection.
Verifying this property was only a small extension to the rest of the verification.
(For simplicity, Figures \ref{fig:exampleDefs}-\ref{fig:exampleAlloc} omit this
property, but the practical garbage collectors in the public source release
include verification of this property.)

\subsection{Allocation, marking, and sweeping}

\begin{figure}

\begin{flushleft}

\verb``

\verb`function{:expand false} T(i) { true }`

\verb`const NO_ABS:int, memLo:int, memHi:int;`

\verb`axiom 0 < memLo` \verb`<=` \verb`memHi;`

\verb`fun memAddr(i) { memLo` \verb`<=` \verb`i` \verb`<` \verb`memHi }`

\verb``

\verb`fun Unalloc(i) { i` \verb`==` \verb`0 }`

\verb`fun White(i) { i` \verb`==` \verb`1 }`

\verb`fun Gray(i) { i` \verb`==` \verb`2 }`

\verb`fun Black(i) { i` \verb`==` \verb`3 }`

\verb``

\verb`var Mem:[int,int]int, Color:[int]int;`

\verb`var $toAbs:[int]int, $AbsMem:[int,int]int;`

\verb``

\verb`fun WellFormed($toAbs) {`

\verb`  (`$\forall^{\tt T}\!$\verb`i1`$.\!$\verb``$\forall^{\tt T}\!$\verb`i2`$.$

\verb`        memAddr(i1) && memAddr(i2)`

\verb`     && $toAbs[i1]` \verb`!=` \verb`NO_ABS`

\verb`     && $toAbs[i2]` \verb`!=` \verb`NO_ABS`

\verb`     && i1` \verb`!=` \verb`i2`

\verb`  ` \verb`==>` \verb`$toAbs[i1]` \verb`!=` \verb`$toAbs[i2])`

\verb`}`

\verb`fun Pointer($toAbs`, \verb`ptr`, \verb`$abs) {`

\verb`    memAddr(ptr) && $abs` \verb`!=` \verb`NO_ABS && $toAbs[ptr]` \verb`==` \verb`$abs`

\verb`}`

\verb`fun ObjInv(i`, \verb`$toAbs`, \verb`$AbsMem`, \verb`Mem) {`

\verb`  $toAbs[i]` \verb`!=` \verb`NO_ABS ==>`

\verb`      Pointer($toAbs`, \verb`Mem[i,0]`, \verb`$AbsMem[$toAbs[i],0])`

\verb`   && Pointer($toAbs`, \verb`Mem[i,1]`, \verb`$AbsMem[$toAbs[i],1])`

\verb`}`

\verb`fun GcInv(Color`, \verb`$toAbs`, \verb`$AbsMem`, \verb`Mem) {`

\verb`    WellFormed($toAbs)`

\verb` && (`$\forall^{\tt T}\!$\verb`i`$.\!$\verb` memAddr(i) ==>`

\verb`            ObjInv(i`, \verb`$toAbs`, \verb`$AbsMem`, \verb`Mem)`

\verb`         && 0` \verb`<=` \verb`Color[i]` \verb`<` \verb`4`

\verb`         && (Black(Color[i])` \verb`==>` \verb`!White(Color[Mem[i,0]])`

\verb`                             && !White(Color[Mem[i,1]]))`

\verb`         && ($toAbs[i]` \verb`==` \verb`NO_ABS` \verb`<==>` \verb`Unalloc(Color[i])))`

\verb`}`

\verb`fun MutatorInv(Color`, \verb`$toAbs`, \verb`$AbsMem`, \verb`Mem) {`

\verb`    WellFormed($toAbs)`

\verb` && (`$\forall^{\tt T}\!$\verb`i`$.\!$\verb` memAddr(i) ==>`

\verb`            ObjInv(i`, \verb`$toAbs`, \verb`$AbsMem`, \verb`Mem)`

\verb`         && 0` \verb`<=` \verb`Color[i]` \verb`<` \verb`2`

\verb`         && ($toAbs[i]` \verb`==` \verb`NO_ABS` \verb`<==>` \verb`Unalloc(Color[i])))`

\verb`}`

\end{flushleft}
\caption{\small \it \textbf{Miniature Mark-Sweep Collector: Definitions.}}
\label{fig:exampleDefs}
\end{figure}

\begin{figure}
\begin{flushleft}
\verb``

\verb`procedure Initialize()`

\verb`  modifies $toAbs`, \verb`Color;`

\verb`  ensures  MutatorInv(Color`, \verb`$toAbs`, \verb`$AbsMem`, \verb`Mem);`

\verb`  ensures  WellFormed($toAbs);`

\verb`{`

\verb`  var ptr;`

\verb`  ptr := memLo;`

\verb`  while (ptr` \verb`<` \verb`memHi)`

\verb`    invariant T(ptr) && memLo` \verb`<=` \verb`ptr` \verb`<=` \verb`memHi;`

\verb`    invariant (`$\forall^{\tt T}\!$\verb`i`$.\!$\verb` memLo` \verb`<=` \verb`i <ptr ==>`

\verb`                   $toAbs[i]` \verb`==` \verb`NO_ABS && Unalloc(Color[i]));`

\verb`  {`

\verb`    Color[ptr] := 0;`

\verb`    $toAbs[ptr] := NO_ABS;`

\verb`    ptr := ptr + 1;`

\verb`  }`

\verb`}`

\verb``

\verb``

\verb`procedure ReadField(ptr`, \verb`field) returns (val)`

\verb`  requires MutatorInv(Color`, \verb`$toAbs`, \verb`$AbsMem`, \verb`Mem);`

\verb`  requires Pointer($toAbs`, \verb`ptr`, \verb`$toAbs[ptr]);`

\verb`  requires field` \verb`==` \verb`0 || field` \verb`==` \verb`1;`

\verb`  ensures  Pointer($toAbs`, \verb`val`, \verb`$AbsMem[$toAbs[ptr],field]);`

\verb`{`

\verb`  assert T(ptr);`

\verb`  val := Mem[ptr,field];`

\verb`}`

\verb``

\verb``

\verb`procedure WriteField(ptr`, \verb`field`, \verb`val)`

\verb`  requires MutatorInv(Color`, \verb`$toAbs`, \verb`$AbsMem`, \verb`Mem);`

\verb`  requires Pointer($toAbs`, \verb`ptr`, \verb`$toAbs[ptr]);`

\verb`  requires Pointer($toAbs`, \verb`val`, \verb`$toAbs[val]);`

\verb`  requires field` \verb`==` \verb`0 || field` \verb`==` \verb`1;`

\verb`  modifies $AbsMem`, \verb`Mem;`

\verb`  ensures  MutatorInv(Color`, \verb`$toAbs`, \verb`$AbsMem`, \verb`Mem);`

\verb`  ensures  $AbsMem == old($AbsMem)[$toAbs[ptr],field := $toAbs[val]];`

\verb`{`

\verb`  assert T(ptr) && T(val);`

\verb`  Mem[ptr,field] := val;`

\verb`  $AbsMem[$toAbs[ptr],field] := $toAbs[val];`

\verb`}`
\end{flushleft}
\caption{\small \it \textbf{Miniature Mark-Sweep Collector: Initialize, ReadField, WriteField.}}
\label{fig:exampleInit}
\end{figure}

\begin{figure}
\begin{flushleft}
\verb`procedure GarbageCollect(root)`

\verb`  requires MutatorInv(Color`, \verb`$toAbs`, \verb`$AbsMem`, \verb`Mem);`

\verb`  requires root` \verb`!=` \verb`0 ==> Pointer($toAbs`, \verb`root`, \verb`$toAbs[root]);`

\verb`  modifies Color`, \verb`$toAbs;`

\verb`  ensures  MutatorInv(Color`, \verb`$toAbs`, \verb`$AbsMem`, \verb`Mem);`

\verb`  ensures  root` \verb`!=` \verb`0 ==> Pointer($toAbs`, \verb`root`, \verb`$toAbs[root]);`

\verb`  ensures  `$\!$\verb`(`$\forall^{\tt T}\!$\verb`i`$.\!\!\!$\verb` memAddr(i)` \verb`&& $toAbs[i]` \verb`!=` \verb`NO_ABS ==>`

\verb`               $toAbs[i]` \verb`==` \verb`old($toAbs)[i]);`

\verb`  ensures  root` \verb`!=` \verb`0 ==> $toAbs[root] == old($toAbs)[root];`

\verb`{`

\verb`  assert T(root);`

\verb`  if (root` \verb`!=` \verb`0)`

\verb`  {`

\verb`    call Mark(root);`

\verb`  }`

\verb`  call Sweep();`

\verb`}`

\verb``

\verb``

\verb``

\verb``

\verb``

\verb``

\verb``

\verb`procedure Mark(ptr)`

\verb`  requires GcInv(Color`, \verb`$toAbs`, \verb`$AbsMem`, \verb`Mem);`

\verb`  requires memAddr(ptr) && T(ptr);`

\verb`  requires $toAbs[ptr]` \verb`!=` \verb`NO_ABS;`

\verb`  modifies Color;`

\verb`  ensures  GcInv(Color`, \verb`$toAbs`, \verb`$AbsMem`, \verb`Mem);`

\verb`  ensures  (`$\forall^{\tt T}\!$\verb`i`$.\!$\verb` !Black(Color[i]) ==> Color[i]` \verb`==` \verb`old(Color)[i]);`

\verb`  ensures  !White(Color[ptr]);`

\verb`{`

\verb`  if (White(Color[ptr]))`

\verb`  {`

\verb`    Color[ptr] := 2; // make gray`

\verb`    call Mark(Mem[ptr,0]);`

\verb`    call Mark(Mem[ptr,1]);`

\verb`    Color[ptr] := 3; // make black`

\verb`  }`

\verb`}`
\end{flushleft}
\caption{\small \it \textbf{Miniature Mark-Sweep Collector: Garbage Collect, Mark.}}
\label{fig:exampleMark}
\end{figure}

\begin{figure}
\begin{flushleft}
\verb`procedure Sweep()`

\verb`  requires GcInv(Color`, \verb`$toAbs`, \verb`$AbsMem`, \verb`Mem);`

\verb`  requires (`$\forall^{\tt T}\!$\verb`i`$.\!$\verb` memAddr(i)` \verb`==>` \verb`!Gray(Color[i]));`

\verb`  modifies Color`, \verb`$toAbs;`

\verb`  ensures  MutatorInv(Color`, \verb`$toAbs`, \verb`$AbsMem`, \verb`Mem);`

\verb`  ensures  (`$\forall^{\tt T}\!$\verb`i`$.\!$\verb` memAddr(i) ==>`

\verb`                    (Black(old(Color)[i])` \verb`==>` \verb`$toAbs[i]` \verb`!=` \verb`NO_ABS)`

\verb`                 && ($toAbs[i]` \verb`!=` \verb`NO_ABS ==> $toAbs[i]` \verb`==` \verb`old($toAbs)[i]));`

\verb`{`

\verb`  var ptr;`

\verb`  ptr := memLo;`

\verb``

\verb`  while (ptr` \verb`<` \verb`memHi)`

\verb`    invariant T(ptr) && memLo` \verb`<=` \verb`ptr` \verb`<=` \verb`memHi;`

\verb`    invariant WellFormed($toAbs);`

\verb`    invariant (`$\forall^{\tt T}\!$\verb`i`$.\!$\verb` memAddr(i) ==>`

\verb`        0` \verb`<=` \verb`Color[i]` \verb`<` \verb`4`

\verb`     && !Gray(Color[i])`

\verb`     && (Black(old(Color)[i]) ==>`

\verb`            $toAbs[i]` \verb`!=` \verb`NO_ABS`

\verb`         && ObjInv(i`, \verb`$toAbs`, \verb`$AbsMem`, \verb`Mem)`

\verb`         && (Mem[i,0]` \verb`>=` \verb`ptr ==> !White(Color[Mem[i,0]]))`

\verb`         && (Mem[i,1]` \verb`>=` \verb`ptr ==> !White(Color[Mem[i,1]])))`

\verb`     && ($toAbs[i]` \verb`==` \verb`NO_ABS` \verb`<==>` \verb`Unalloc(Color[i]))`

\verb`     && ($toAbs[i]` \verb`!=` \verb`NO_ABS ==> $toAbs[i]` \verb`==` \verb`old($toAbs)[i])`

\verb`     && (ptr` \verb`<=` \verb`i` \verb`==>` \verb`Color[i]` \verb`==` \verb`old(Color)[i])`

\verb`     && (i` \verb`<` \verb`ptr` \verb`==>` \verb`0` \verb`<=` \verb`Color[i]` \verb`<` \verb`2)`

\verb`     && (i` \verb`<` \verb`ptr && White(Color[i]) ==> Black(old(Color)[i])));`

\verb`  {`

\verb`    if (White(Color[ptr]))`

\verb`    {`

\verb`      Color[ptr] := 0; // deallocate`

\verb`      $toAbs[ptr] := NO_ABS;`

\verb`    }`

\verb`    else if (Black(Color[ptr]))`

\verb`    {`

\verb`      Color[ptr] := 1; // make white`

\verb`    }`

\verb`    ptr := ptr + 1;`

\verb`  }`

\verb`}`
\end{flushleft}
\caption{\small \it \textbf{Miniature Mark-Sweep Collector: Sweep.}}
\label{fig:exampleSweep}
\end{figure}

\begin{figure}
\begin{flushleft}

\verb``

\verb`procedure Alloc(root`, \verb`$abs) returns (newRoot,ptr)`

\verb`  requires MutatorInv(Color`, \verb`$toAbs`, \verb`$AbsMem`, \verb`Mem);`

\verb`  requires root` \verb`!=` \verb`0 ==> Pointer($toAbs`, \verb`root`, \verb`$toAbs[root]);`

\verb`  requires $abs` \verb`!=` \verb`NO_ABS;`

\verb`  requires (`$\forall^{\tt T}\!$\verb`i`$.\!$\verb` memAddr(i)` \verb`==>` \verb`$toAbs[i]` \verb`!=` \verb`$abs);`

\verb`  requires $AbsMem[$abs,0]` \verb`==` \verb`$abs;`

\verb`  requires $AbsMem[$abs,1]` \verb`==` \verb`$abs;`

\verb`  modifies Color`, \verb`$toAbs`, \verb`Mem;`

\verb`  ensures  MutatorInv(Color`, \verb`$toAbs`, \verb`$AbsMem`, \verb`Mem);`

\verb`  ensures  root` \verb`!=` \verb`0 ==> Pointer($toAbs`,\verb`newRoot`,\verb`old($toAbs)[root]);`

\verb`  ensures  Pointer($toAbs`, \verb`ptr`, \verb`$abs);`

\verb`  ensures  WellFormed($toAbs);`

\verb`{`

\verb``

\verb`  while (true)`

\verb`    invariant MutatorInv(Color`, \verb`$toAbs`, \verb`$AbsMem`, \verb`Mem);`

\verb`    invariant root` \verb`!=` \verb`0 ==> Pointer($toAbs`, \verb`root`, \verb`$toAbs[root]);`

\verb`    invariant (`$\forall^{\tt T}\!$\verb`i`$.\!$\verb` memAddr(i)` \verb`==>` \verb`$toAbs[i]` \verb`!=` \verb`$abs);`

\verb`    invariant root != 0 ==> $toAbs[root] == old($toAbs)[root];`

\verb``

\verb`  {`

\verb`    ptr := memLo;`

\verb``

\verb`    while (ptr` \verb`<` \verb`memHi)`

\verb`      invariant T(ptr) && memLo` \verb`<=` \verb`ptr` \verb`<=` \verb`memHi;`

\verb`    {`

\verb`      if (Unalloc(Color[ptr]))`

\verb`      {`

\verb`        Color[ptr] := 1; // make white`

\verb`        $toAbs[ptr] := $abs;`

\verb`        Mem[ptr,0] := ptr;`

\verb`        Mem[ptr,1] := ptr;`

\verb`        newRoot := root;`

\verb`        return;`

\verb`      }`

\verb`      ptr := ptr + 1;`

\verb`    }`

\verb``

\verb`    call GarbageCollect(root);`

\verb`  }`

\verb`}`
\end{flushleft}
\caption{\small \it \textbf{Miniature Mark-Sweep Collector: Allocation.}}
\label{fig:exampleAlloc}
\end{figure}

Figure \ref{fig:exampleAlloc}'s \verb`Alloc` procedure performs an (inefficient) linear search for a free memory address; if no free space remains, \verb`Alloc` calls the garbage collector.  The collector recursively marks all nodes reachable from some
root pointer (the ``mark phase''), and then deallocates all unmarked
objects (the ``sweep phase'').  Figures \ref{fig:exampleMark} and \ref{fig:exampleSweep} show the code for
the \verb`Mark` and \verb`Sweep` procedures.  The next few paragraphs trace
the preconditions and postconditions for \verb`Mark` and \verb`Sweep` backwards, starting with \verb`Sweep`'s postconditions.

A key property of \verb`Sweep` is that it leaves no dangling pointers (pointers from allocated objects to free space).  This property is part of \verb`MutatorInv`:  each memory address \verb`i` satisfies \verb`ObjInv(i, ...)`, which ensures that if some object lives at \verb`i` (if \verb`$toAbs[i] != NO_ABS`), then the object's fields contain valid pointers to allocated objects (see Figure \ref{fig:exampleDefs}).  Specifically, the fields \verb`Mem[i,0]` and \verb`Mem[i,1]` are, like \verb`i`, mapped to some abstract nodes, so that \verb`$toAbs[Mem[i,0]] != NO_ABS` $\!$ and \verb`$toAbs[Mem[i,1]] != NO_ABS`.  To maintain this property, \verb`Sweep` must ensure that any object it deallocates had no pointers from objects that remain allocated.  Since \verb`Sweep` deallocates white objects and leaves gray and black objects allocated, \verb`Sweep`'s preconditions requires that no gray-to-white or black-to-white pointers exist.

To rule out gray-to-white pointers, \verb`Sweep`'s second precondition requires that no gray objects exist at all:

\begin{flushleft}
\[
\texttt{requires (}\forall^{\tt T}\!\texttt{i}.\!\texttt{ memAddr(i)}\texttt{ ==> }\texttt{!Gray(Color[i]));}
\]
\end{flushleft}
The \verb`GcInv` function (see Figure \ref{fig:exampleDefs}) prohibits black-to-white
pointers: every black object has fields pointing to non-white
objects.  (This is known as the tri-color or three color invariant
\cite{dijk76b}.)

The \verb`Mark` procedure's postconditions must satisfy \verb`Sweep`'s preconditions.
To ensure that no gray objects exist at the end of the mark phase,
\verb`Mark`'s second postcondition says that any non-black object at the end of the mark phase
retained its original color from the beginning of the mark phase.
For example, any leftover gray objects must have been gray at the beginning of the mark phase.
Since no gray objects existed at the beginning, no gray objects exist at the end.
\verb`Mark` obeys the ban on black-to-white pointers by coloring an object black {\em after} its children are non-white.
(Before coloring a node's children, \verb`Mark` temporarily colors the node gray to indicate the node is ``in progress'';
without this intermediate step, a cycle in the graph would send \verb`Mark` into an infinite loop.)

\subsection{Quantifiers and triggers}\label{subsec-trigger}

In the absence of universal and existential quantifiers, many
theories are decidable and have practical decision procedures.
These include the theory of arrays, the theory of linear
arithmetic, the theory of uninterpreted functions, and the
combination of these theories.  Unfortunately, adding quantifiers
makes the theories either undecidable or very slow to decide: the
combination of linear arithmetic and arrays, for example, is
undecidable in the presence of quantifiers. 
This forces
verification to rely on heuristics for instantiating
quantifiers.  The choice of heuristics determines the success of
the verification.

Many automated theorem provers,
including Z3~\cite{Z3:Efficient,Z3:Triggers} and Simplify~\cite{Simplify:JACM},
use programmer-supplied {\em triggers} to guide
quantifier instantiation.
(Many other automated theorem provers, such as CVC3~\cite{CVC3:Quantified} and Yices~\cite{Yices:Tool},
use triggers internally, but do not expose triggers directly to programmers.)
Consider again \verb`Sweep`'s precondition
prohibiting gray objects.  Here are two ways to write this in
BoogiePL syntax, each with a different trigger:
\begin{verbatim}
  forall i::{memAddr(i)}memAddr(i)==>!Gray(Color[i]))
  forall i::{Color[i]}  memAddr(i)==>!Gray(Color[i]))
\end{verbatim}
Both have the same logical meaning, but use different instantiation
strategies.  The first asks \verb`i` to be instantiated with
expression \verb`e` whenever an expression \verb`memAddr(e)` appears
during an attempt to prove a theorem. The second asks \verb`i` to be
instantiated with \verb`e` whenever \verb`Color[e]` appears. Selecting
appropriate triggers is challenging in general.  With an overly
selective trigger, a quantified formula may never get instantiated,
leaving a theorem unproved.  With an overly liberal trigger, a
quantified formula may be instantiated too often (even infinitely
often), drowning the theorem prover in unwanted information.

Shaz Qadeer suggested that we look at formulas of form
\verb`forall i::{f(i)}f(i) ==> P`, using \verb`f(i)` as a trigger.  For example,
we could use \verb`memAddr(i)` as a trigger, although this appears
in so many places that it would be easy to accidentally introduce an infinite
instantiation loop.  (The appearance of \verb`memAddr(ptr)` inside
the \verb`Pointer` function, which in turn appears in the
\verb`ObjInv` function, which in turn appears in the \verb`GcInv`
function, is one example of such a loop.)  To avoid accidental
loops, we introduce a function \verb`T(i:int)`,
solely for use as a trigger, writing the
invariants above as:

\begin{verbatim}
  forall i::{T(i)}T(i)==>memAddr(i)==>!Gray(Color[i]))
\end{verbatim}
(Note that the \verb`==>` operator is right associative.)

We define the function \verb`T` to be true everywhere: for all \verb`i`,
\verb`T(i) == true`.  Thus, adding \verb`T(e)` to a logical formula
doesn't change the purely logical meaning of the formula.
However, \verb`T(e)` does function as a hint to Z3, indicating that \verb`e`
is an interesting expression that should be used to instantiate quantifiers.
In this way, adding instances of \verb`T(e)` for various \verb`e` can
guide Z3's quantifier instantiation, as illustrated further below.

For conciseness, we abbreviate \verb`forall i::{T(i)}T(i)==>` as $\forall^{\tt T}\texttt{i}$.
To avoid instantiation loops, we never write a formula of the form $\forall^{\tt T}\texttt{i}.(...\texttt{T(e)}...)$, where \verb`e` is some expression other than a simple quantified variable.

Based on the trigger \verb`T(i)`, we use two strategies to ensure sufficient instantiation of quantified formulas.
First, we write explicit assertions of \verb`T(e)` for various expressions \verb`e` that appear in the program.
This helps Z3 prove formulas $\texttt{(}\forall^{\tt T}\texttt{i.P(i))==>P(e)}$.
For example, the \verb`ReadField` procedure explicitly asserts \verb`T(ptr)` to instantiate the quantifiers in \verb`MutatorInv` at the value \verb`ptr`.

Second, we use the trigger \verb`T(i)` to prove formulas of the form $\texttt{(}\forall^{\tt T}\texttt{i.P(i))==>(}\forall^{\tt T}\texttt{j.Q(j))}$.
In this case, since \verb`T` appears in both quantifiers, Z3 automatically instantiates \verb`P` at \verb`i=j` to prove \verb`Q(j)`.
This second strategy isn't sufficient for all \verb`P` and \verb`Q`; for example, knowing $\forall^{\tt T}\texttt{i.a[i + 5] == 0}$ does not prove $\forall^{\tt T}\texttt{j.a[j + 6] == 0}$, even though mathematically, both these formulas are equivalent.
Nevertheless, this strategy works well for purely local reasoning.
For example, \verb`Sweep`'s loop invariant maintains the property:
\[
\forall^{\tt T}\texttt{i.memAddr(i)==>!Gray(Color[i])}
\]
If the loop updates \verb`Color` by changing \verb`Color[ptr]` to \verb`1` (white), then the theorem prover attempts to prove:

\begin{verbatim}
      (memAddr(i)==>!Gray(Color[i]))
  ==> (memAddr(i)==>!Gray(Color'[i]))
\end{verbatim}
where \verb`Color' == Color[ptr := 1]`. In the case where \verb`i != ptr`, \verb`Color[i] == Color'[i]` and the proof is trivial. In the case where \verb`i == ptr`, \verb`!Gray(Color'[i]) == !Gray(1) == true`. The proof is easy because the formula \verb`memAddr(i) ==>`$\ \ \ \ \ \ $ \verb`!Gray(Color[i])` is entirely local; it depends only on array elements at index \verb`i`.

Many formulas depend on non-local array elements, though. Consider how \verb`Mark` maintains this piece of the tri-color invariant (no black-to-white pointers) from \verb`GcInv` in Figure \ref{fig:exampleDefs}:

\begin{verbatim}
  Black(Color[i]) ==> !White(Color[Mem[i,0]])
\end{verbatim}
This depends not only on \verb`i`'s color, but on the color of some other node \verb`Mem[i,0]`.
For non-local formulas, the local instantiation strategy suffices for some programs but not for others.
For example, it suffices for the collector in Figures \ref{fig:exampleDefs}-\ref{fig:exampleAlloc}
(we invite the reader to write out the verification conditions by hand to see),
but did not suffice for an analogous copying collector that we wrote
(it did not sufficiently instantiate information about objects pointed to by forwarding pointers).
This limitation motivated the use of regions, as described in the next section.

\section{Regions}\label{sec-regions}

A mark-sweep collector appears easier to verify than a copying collector, because the mark-sweep collector doesn't modify pointers inside objects.  As the previous section mentioned, the mark-sweep collector in Figures \ref{fig:exampleDefs}-\ref{fig:exampleAlloc} passed verification even with a very simple triggering strategy, while the analogous copying collector did not.  Therefore, this section augments the two strategies described in the previous section with a third instantiation strategy, based on {\em regions}.
Together, these three strategies were sufficient for both mark-sweep and copying collectors.
(Although regions aren't {\em necessary} for our mark-sweep collector,
and can be omitted for strictly non-moving mark-sweep collectors,
regions would be useful for mark-sweep collectors that employ compaction,
or for collectors that combine mark-sweep and copying collection.)

Regions have proven useful for verifying the type safety of copying collectors \cite{wang01,monn01}, which suggests that they might also help verify the {\em correctness} of copying collectors.
Type systems for regions
are similar to the verification presented in Section \ref{sec-miniature-ms}: Section
\ref{sec-miniature-ms}'s verification mapped concrete addresses to abstract nodes,
while type systems type-check a region by mapping concrete
addresses in the region to types (e.g., a type system with
types \verb`Parent` and \verb`Child` might map Figure \ref{fig:conabs}'s \verb`C1` to \verb`Parent` and \verb`C2` and
\verb`C3` to \verb`Child`). This suggests a strategy for importing regions (and
the ease of verifying copying collectors via regions) from type
systems: rather than defining just one concrete-to-abstract
mapping \verb`$toAbs`, allow multiple regions, where each region
is an independent concrete-to-abstract mapping.

For example, consider how Figure \ref{fig:exampleDefs}'s object invariant uses \verb`$toAbs`:
\begin{verbatim}
ObjInv(i,$toAbs,$AbsMem,Mem) =
  $toAbs[i] != NO_ABS ==>
     Pointer($toAbs, Mem[i,0], $AbsMem[$toAbs[i],0])
     ...
\end{verbatim}
Expanding the \verb`Pointer` function exposes a non-local invariant:
\begin{verbatim}
  $toAbs[i] != NO_ABS ==>
     ... $toAbs[Mem[i,0]] != NO_ABS ...
\end{verbatim}
This invariant is crucial; as discussed in Section \ref{sec-miniature-ms}, it ensures that no dangling pointers exist.  However, it's not obvious how to prove that this invariant is maintained when \verb`$toAbs[Mem[i,0]]` changes.  Therefore, the remainder of this paper adopts a region-based object invariant:
\begin{verbatim}
  ObjInv(i,$rs,$rt,$toAbs,$AbsMem,Mem) =
    $rs[i] != NO_ABS ==>
       Pointer($rt, Mem[i,0], $AbsMem[$toAbs[i],0])
   ...
\end{verbatim}
This object invariant describes an object living in a source region \verb`$rs`, whose fields point to some target region \verb`$rt`.  Expanding the Pointer function yields:
\begin{verbatim}
  $rs[i] != NO_ABS ==>
     ... $rt[Mem[i,0]] != NO_ABS ...
\end{verbatim}
Now we adopt another idea from region-based type systems: regions only grow over time, and are then deallocated all at once; deallocating a single object from a region is not allowed.  In our setting, this means that for any address \verb`j` and region \verb`$r`, \verb`$r[j]` may change monotonically from \verb`NO_ABS` to some particular abstract node, but thereafter \verb`$r[j]` is fixed at that abstract node.  The function \verb`RExtend` expresses this restriction; the memory manager only changes \verb`$r` to some new \verb`$r'` if \verb`RExtend($r,$r')` holds:
\begin{verbatim}
  fun RExtend($r:[int]int,$r':[int]int) {
    (forall i::{$r[i]}{$r'[i]}
      $r[i] != NO_ABS ==> $r[i] == $r'[i])
  }
\end{verbatim}
RExtend's quantifier is {\em not} based on \verb`T`; instead, it can trigger on either \verb`$r[i]` or \verb`$r'[i]`.  (Note that \verb`RExtend` introduces no instantiation loops, because it only mentions \verb`r` and \verb`r'` at index \verb`i`, and does not mention \verb`T` at all.)  In combination with the second strategy from Section \ref{sec-miniature-ms}, this triggering allows Z3 to prove formulas of the form $\texttt{(}\forall^{\tt T}\texttt{i.P(r[e]))==>(}\forall^{\tt T}\texttt{i.P(r'[e]))}$, where \verb`e` depends on \verb`i`.
For example, given the guarantee that \verb`RExtend($rt,$rt')`, the object invariant ensures that if \verb`$rt[Mem[i,0]] != NO_ABS`, then \verb`$rt'[Mem[i,0]] != NO_ABS`.

Given this region-based object invariant,
a memory manager can express all other invariants about node \verb`i` as purely local invariants.
For example, our region-based mark-sweep collector relates
\verb`i`'s color to \verb`i`'s region state using purely local reasoning,
using a first region \verb`$r1` to represent the set of all currently allocated objects
and a second region \verb`$r2` to represent the set of objects reached so far
during the current collection:
\begin{verbatim}
     (White(Color[i]) ==>
          $r1[i] != NO_ABS && $r2[i] == NO_ABS
       && ObjInv(i,$r1,$r1,$toAbs,$AbsMem,Mem))
  && (Gray(Color[i]) ==>
          $r1[i] != NO_ABS && $r2[i] != NO_ABS
       && $r1[i] == $r2[i]
       && ObjInv(i,$r1,$r1,$toAbs,$AbsMem,Mem))
  && (Black(Color[i]) ==>
          $r1[i] != NO_ABS && $r2[i] != NO_ABS
       && $r1[i] == $r2[i]
       && ObjInv(i,$r2,$r2,$toAbs,$AbsMem,Mem))
\end{verbatim}
If \verb`i` is black, then \verb`ObjInv(i,$r2,$r2,...)` ensures that
\verb`i`'s fields point to members of region \verb`$r2`.  Members of
\verb`$r2` cannot be white, since the invariant above forces white
nodes to {\em not} be members of \verb`$r2`.  Thus, the invariant
indirectly expresses the standard tri-color invariant (no
black-to-white pointers), and the collector need not state the
tri-color invariant directly.

We briefly sketch the region lifetimes during a mark-sweep garbage collection.  The collector's mark phase begins with \verb`$r1` equal to \verb`$toAbs` and \verb`$r2` empty (i.e. \verb`$r2` maps all nodes to \verb`NO_ABS`).  At the beginning of the mark phase, all allocated objects are white, so the invariant above needs \verb` ObjInv(i,$r1,$r1,...)`, and requires that no objects be members of \verb`$r2`.  As the mark phase marks each reached node \verb`i` gray, it adds \verb`i` to \verb`$r2`, so that \verb`$r2[i] != NO_ABS`.
At the end of the mark phase, \verb`$r2` contains exactly the reached objects,
while \verb`$r1` and \verb`$toAbs` are the same as at the beginning of the mark phase.
The sweep phase then removes unreached objects from \verb`$toAbs` until \verb`$toAbs == $r2`;
\verb`Sweep` leaves \verb`$r1` and \verb`$r2` unmodified.
After sweeping, only the objects in \verb`$r2` remain allocated
(sweeping removes all objects in \verb`$r1` that aren't in \verb`$r2`).
At this point, \verb`$r1` is no longer useful,
so the mutator takes an action analogous to ``deallocating'' region \verb`$r1`:
it simply forgets about \verb`$r1`,
throwing out all invariants relating to \verb`$r1` and keeping only the invariants for \verb`$r2`.
In the next collection cycle, \verb`$r2` becomes the new \verb`$r1`, and the process repeats.

\begin{figure}
\begin{flushleft}
\verb`function{:expand false} T(i) { true }`

\verb`const NO_ABS:int`, \verb`memLo:int`, \verb`memMid:int`, \verb`memHi:int;`

\verb`const MAP_NO_ABS:[int]int;`

\verb`axiom (`$\forall^{\tt T}\!$\verb`i`$.\!$\verb` MAP_NO_ABS[i]` \verb`==` \verb`NO_ABS);`

\verb`axiom 0` \verb`<` \verb`memLo && memLo` \verb`<=` \verb`memMid && memMid` \verb`<=` \verb`memHi;`

\verb`fun memAddr(i) { memLo` \verb`<=` \verb`i` \verb`<` \verb`memHi }`

\verb``

\verb`var Mem:[int,int]int`, \verb`FwdPtr:[int]int;`

\verb`var $toAbs:[int]int`, \verb`$AbsMem:[int,int]int;`

\verb`var $r1:[int]int`, \verb`$r2:[int]int;`

\verb``

\verb`// Fromspace ranges from Fi to Fl`, \verb`where Fk..Fl is empty`

\verb`//   Tospace ranges from Ti to Tl`, \verb`where Tk..Tl is empty`

\verb`var Fi:int;`

\verb`var Fk:int;`

\verb`var Fl:int;`

\verb`var Ti:int;`

\verb`var Tj:int;`

\verb`var Tk:int;`

\verb`var Tl:int;`

\verb``

\verb`fun WellFormed($r) {`

\verb`  (`$\forall^{\tt T}\!$\verb`i1`$.\!$\verb``$\forall^{\tt T}\!$\verb`i2`$.\!$\verb` memAddr(i1)`

\verb`     && memAddr(i2)`

\verb`     && $r[i1]` \verb`!=` \verb`NO_ABS`

\verb`     && $r[i2]` \verb`!=` \verb`NO_ABS`

\verb`     && i1` \verb`!=` \verb`i2`

\verb`  ` \verb`==>` \verb`$r[i1]` \verb`!=` \verb`$r[i2])`

\verb`}`

\verb``

\verb`fun Pointer($r`, \verb`ptr`, \verb`$abs) {`

\verb`    memAddr(ptr) && $abs` \verb`!=` \verb`NO_ABS`

\verb` && $r[ptr]` \verb`==` \verb`$abs`

\verb`}`

\verb``

\verb`fun ObjInv(i`, \verb`$rs`, \verb`$rt`, \verb`$toAbs`, \verb`$AbsMem`, \verb`Mem) {`

\verb`  $rs[i]` \verb`!=` \verb`NO_ABS ==>`

\verb`      Pointer($rt`, \verb`Mem[i,0]`, \verb`$AbsMem[$toAbs[i],0])`

\verb`   && Pointer($rt`, \verb`Mem[i,1]`, \verb`$AbsMem[$toAbs[i],1])`

\verb`}`

\verb``
\end{flushleft}
\caption{\small \it \textbf{Miniature Copying Collector: Definitions.}}
\label{fig:exampleCopyDef}
\end{figure}

\begin{figure}
\begin{flushleft}
\verb`fun GcInv(FwdPtr`, \verb`Fi`, \verb`Fk`, \verb`Fl`, \verb`Ti`, \verb`Tj`, \verb`Tk`, \verb`Tl,`

\verb`               $r1`, \verb`$r2`, \verb`$toAbs`, \verb`$AbsMem`, \verb`Mem) {`

\verb`    WellFormed($toAbs)`

\verb` && memLo` \verb`<=` \verb`Fi && Fi` \verb`<=` \verb`Fk && Fk` \verb`<=` \verb`Fl && Fl` \verb`<=` \verb`memHi`

\verb` && memLo` \verb`<=` \verb`Ti && Ti` \verb`<=` \verb`Tj && Tj` \verb`<=` \verb`Tk && Tk` \verb`<=` \verb`Tl && Tl` \verb`<=` \verb`memHi`

\verb` && (Fl` \verb`<=` \verb`Ti || Tl` \verb`<=` \verb`Fi)`

\verb` && (`$\forall^{\tt T}\!$\verb`i`$.\!$\verb` memAddr(i) ==>`

\verb`        ($r2[i]` \verb`!=` \verb`NO_ABS` \verb`==>` \verb`$toAbs[i]` \verb`==` \verb`$r2[i])`

\verb`     && ($r1[i]` \verb`!=` \verb`NO_ABS` \verb`<==>` \verb`Fi` \verb`<=` \verb`i` \verb`<` \verb`Fk)`

\verb`     && ($r2[i]` \verb`!=` \verb`NO_ABS` \verb`<==>` \verb`Ti` \verb`<=` \verb`i` \verb`<` \verb`Tk)`

\verb`     && (Fi` \verb`<=` \verb`i` \verb`<` \verb`Fk ==>`

\verb`            (FwdPtr[i]` \verb`==` \verb`0` \verb`<==>` \verb`$toAbs[i]` \verb`!=` \verb`NO_ABS)`

\verb`         && (FwdPtr[i]` \verb`!=` \verb`0` \verb`==>` \verb`Pointer($r2`, \verb`FwdPtr[i]`, \verb`$r1[i]))`

\verb`         && (FwdPtr[i]` \verb`==` \verb`0` \verb`==>` \verb`$toAbs[i]` \verb`==` \verb`$r1[i]`

\verb`                             && ObjInv(i`, \verb`$r1`, \verb`$r1`, \verb`$toAbs`, \verb`$AbsMem`, \verb`Mem)))`

\verb`     && (Ti` \verb`<=` \verb`i` \verb`<` \verb`Tk` \verb`==>`

\verb`             FwdPtr[i]` \verb`==` \verb`0 && $toAbs[i]` \verb`!=` \verb`NO_ABS && $toAbs[i]` \verb`==` \verb`$r2[i])`

\verb`     && (Ti` \verb`<=` \verb`i` \verb`<` \verb`Tj` \verb`==>` \verb`ObjInv(i`, \verb`$r2`, \verb`$r2`, \verb`$toAbs`, \verb`$AbsMem`, \verb`Mem))`

\verb`     && (Tj` \verb`<=` \verb`i` \verb`<` \verb`Tk` \verb`==>` \verb`ObjInv(i`, \verb`$r2`, \verb`$r1`, \verb`$toAbs`, \verb`$AbsMem`, \verb`Mem)))`

\verb`}`

\verb``

\verb``

\verb`fun MutatorInv(FwdPtr`, \verb`Fi`, \verb`Fk`, \verb`Fl`, \verb`Ti`, \verb`Tj`, \verb`Tk`, \verb`Tl,`

\verb`                    $toAbs`, \verb`$AbsMem`, \verb`Mem) {`

\verb`    WellFormed($toAbs)`

\verb` && memLo` \verb`<=` \verb`Fi && Fi` \verb`<=` \verb`Fk && Fk` \verb`<=` \verb`Fl && Fl` \verb`<=` \verb`memHi`

\verb` && memLo` \verb`<=` \verb`Ti && Ti` \verb`==` \verb`Tj && Tj` \verb`==` \verb`Tk && Tk` \verb`<=` \verb`Tl && Tl` \verb`<=` \verb`memHi`

\verb` && (Fl` \verb`<=` \verb`Ti || Tl` \verb`<=` \verb`Fi)`

\verb` && (`$\forall^{\tt T}\!$\verb`i`$.\!$\verb` memAddr(i) ==>`

\verb`        ObjInv(i`, \verb`$toAbs`, \verb`$toAbs`, \verb`$toAbs`, \verb`$AbsMem`, \verb`Mem)`

\verb`     && (Fi` \verb`<=` \verb`i` \verb`<` \verb`Fk` \verb`==>` \verb`FwdPtr[i]` \verb`==` \verb`0)`

\verb`     && ($toAbs[i]` \verb`!=` \verb`NO_ABS` \verb`<==>` \verb`Fi` \verb`<=` \verb`i` \verb`<` \verb`Fk))`

\verb`}`

\verb``

\verb``

\verb`// As a region evolves`, \verb`it adds new mappings`, \verb`but each mapping is`

\verb`// permanent: RExtend ensures that new mappings do not overwrite old mappings.`

\verb`fun RExtend(rOld`, \verb`rNew) returns (bool)`

\verb`{`

\verb`  (forall i::{rOld[i]}{rNew[i]} rOld[i]` \verb`!=` \verb`NO_ABS` \verb`==>` \verb`rOld[i]` \verb`==` \verb`rNew[i])`

\verb`}`

\verb``

\end{flushleft}
\caption{\small \it \textbf{Miniature Copying Collector: Definitions, continued.}}
\label{fig:exampleCopyDefs}
\end{figure}

\begin{figure}
\begin{flushleft}
\verb`procedure Initialize()`

\verb`  modifies $toAbs`, \verb`FwdPtr`, \verb`Fi`, \verb`Fk`, \verb`Fl`, \verb`Ti`, \verb`Tj`, \verb`Tk`, \verb`Tl;`

\verb`  ensures  MutatorInv(FwdPtr`, \verb`Fi`, \verb`Fk`, \verb`Fl`, \verb`Ti`, \verb`Tj`, \verb`Tk`, \verb`Tl`, \verb`$toAbs`, \verb`$AbsMem`, \verb`Mem);`

\verb`  ensures  WellFormed($toAbs);`

\verb`{`

\verb`  $toAbs := MAP_NO_ABS;`

\verb`  Fi := memLo;`

\verb`  Fk := memLo;`

\verb`  Fl := memMid;`

\verb`  Ti := memMid;`

\verb`  Tj := memMid;`

\verb`  Tk := memMid;`

\verb`  Tl := memHi;`

\verb`}`

\verb``

\verb``

\verb`procedure ReadField(ptr`, \verb`field) returns (val)`

\verb`  requires MutatorInv(FwdPtr`, \verb`Fi`, \verb`Fk`, \verb`Fl`, \verb`Ti`, \verb`Tj`, \verb`Tk`, \verb`Tl`, \verb`$toAbs`, \verb`$AbsMem`, \verb`Mem);`

\verb`  requires Pointer($toAbs`, \verb`ptr`, \verb`$toAbs[ptr]);`

\verb`  requires field` \verb`==` \verb`0 || field` \verb`==` \verb`1;`

\verb`  ensures  Pointer($toAbs`, \verb`val,`

\verb`                   $AbsMem[$toAbs[ptr],field]);`

\verb`{`

\verb`  assert T(ptr);`

\verb`  val := Mem[ptr,field];`

\verb`}`

\verb``

\verb``

\verb`procedure WriteField(ptr`, \verb`field`, \verb`val)`

\verb`  requires MutatorInv(FwdPtr`, \verb`Fi`, \verb`Fk`, \verb`Fl`, \verb`Ti`, \verb`Tj`, \verb`Tk`, \verb`Tl`, \verb`$toAbs`, \verb`$AbsMem`, \verb`Mem);`

\verb`  requires Pointer($toAbs`, \verb`ptr`, \verb`$toAbs[ptr]);`

\verb`  requires Pointer($toAbs`, \verb`val`, \verb`$toAbs[val]);`

\verb`  requires field` \verb`==` \verb`0 || field` \verb`==` \verb`1;`

\verb`  modifies $AbsMem`, \verb`Mem;`

\verb`  ensures  MutatorInv(FwdPtr`, \verb`Fi`, \verb`Fk`, \verb`Fl`, \verb`Ti`, \verb`Tj`, \verb`Tk`, \verb`Tl`, \verb`$toAbs`, \verb`$AbsMem`, \verb`Mem);`

\verb`  ensures  $AbsMem ==`

\verb`    old($AbsMem)[$toAbs[ptr],field := $toAbs[val]];`

\verb`{`

\verb`  assert T(ptr) && T(val);`

\verb`  Mem[ptr,field] := val;`

\verb`  $AbsMem[$toAbs[ptr],field] := $toAbs[val];`

\verb`}`

\end{flushleft}
\caption{\small \it \textbf{Miniature Copying Collector: Initialization, Read, Write.}}
\label{fig:exampleCopyInitRW}
\end{figure}

\begin{figure}
\begin{flushleft}
\verb``

\verb`procedure forwardFromspacePtr(ptr`, \verb`$freshAbs) returns(ret)`

\verb`  requires GcInv(FwdPtr`, \verb`Fi`, \verb`Fk`, \verb`Fl`, \verb`Ti`, \verb`Tj`, \verb`Tk`, \verb`Tl`, \verb`$r1`, \verb`$r2`, \verb`$toAbs`, \verb`$AbsMem`, \verb`Mem);`

\verb`  requires T(ptr) && Fi` \verb`<=` \verb`ptr` \verb`<` \verb`Fk;`

\verb`  requires T($freshAbs) && $freshAbs` \verb`!=` \verb`NO_ABS;`

\verb`  requires (`$\forall^{\tt T}\!$\verb`i`$.\!$\verb` memAddr(i)` \verb`==>` \verb`$toAbs[i]` \verb`!=` \verb`$freshAbs);`

\verb``

\verb`  modifies FwdPtr`, \verb`$toAbs`, \verb`$r2`, \verb`Tk`, \verb`Mem;`

\verb``

\verb`  ensures  GcInv(FwdPtr`, \verb`Fi`, \verb`Fk`, \verb`Fl`, \verb`Ti`, \verb`Tj`, \verb`Tk`, \verb`Tl`, \verb`$r1`, \verb`$r2`, \verb`$toAbs`, \verb`$AbsMem`, \verb`Mem);`

\verb`  ensures  T(ret) && Pointer($r2`, \verb`ret`, \verb`$r1[ptr]);`

\verb`  ensures  (`$\forall^{\tt T}\!$\verb`i`$.\!$\verb` memAddr(i)` \verb`==>` \verb`$toAbs[i]` \verb`!=` \verb`$freshAbs);`

\verb`  ensures  (`$\forall^{\tt T}\!$\verb`i`$.\!$\verb` i` \verb`!=` \verb`old(Tk)` \verb`==>` \verb`Mem[i`, \verb`0]` \verb`==` \verb`old(Mem)[i`, \verb`0]);`

\verb`  ensures  (`$\forall^{\tt T}\!$\verb`i`$.\!$\verb` i` \verb`!=` \verb`old(Tk)` \verb`==>` \verb`Mem[i`, \verb`1]` \verb`==` \verb`old(Mem)[i`, \verb`1]);`

\verb`  ensures  RExtend(old($r2)`, \verb`$r2);`

\verb`{`

\verb`  if (FwdPtr[ptr]` \verb`!=` \verb`0) {`

\verb``

\verb`    // object already copied`

\verb`    ret := FwdPtr[ptr];`

\verb``

\verb`  }`

\verb`  else {`

\verb`    // copy object to to-space`

\verb``

\verb`    while (Tk` \verb`>=` \verb`Tl) {`

\verb`      // out of memory`

\verb`    }`

\verb``

\verb`    assert T(ptr) && T(Tk);`

\verb`    ret := Tk;`

\verb`    Mem[ret`, \verb`0] := Mem[ptr`, \verb`0];`

\verb`    Mem[ret`, \verb`1] := Mem[ptr`, \verb`1];`

\verb`    FwdPtr[ret] := 0;`

\verb`    $toAbs[ret] := $r1[ptr];`

\verb`    $r2[ret] := $r1[ptr];`

\verb`    $toAbs[ptr] := NO_ABS;`

\verb`    FwdPtr[ptr] := ret;`

\verb`    Tk := Tk + 1;`

\verb``

\verb`  }`

\verb`}`

\verb``

\end{flushleft}
\caption{\small \it \textbf{Miniature Copying Collector: Forwarding.}}
\label{fig:exampleCopyForward}
\end{figure}

\begin{figure}
\begin{flushleft}
\verb`procedure GarbageCollect(root`, \verb`$freshAbs) returns(newRoot)`

\verb`  requires MutatorInv(FwdPtr`, \verb`Fi`, \verb`Fk`, \verb`Fl`, \verb`Ti`, \verb`Tj`, \verb`Tk`, \verb`Tl`, \verb`$toAbs`, \verb`$AbsMem`, \verb`Mem);`

\verb`  requires root` \verb`!=` \verb`0 ==> Pointer($toAbs`, \verb`root`, \verb`$toAbs[root]);`

\verb`  requires T($freshAbs) && $freshAbs` \verb`!=` \verb`NO_ABS;`

\verb`  requires (`$\forall^{\tt T}\!$\verb`i`$.\!$\verb` memAddr(i)` \verb`==>` \verb`$toAbs[i]` \verb`!=` \verb`$freshAbs);`

\verb`  modifies FwdPtr`, \verb`$toAbs`, \verb`$r1`, \verb`$r2`, \verb`Fi`, \verb`Fk`, \verb`Fl`, \verb`Ti`, \verb`Tj`, \verb`Tk`, \verb`Tl`, \verb`Mem;`

\verb`  ensures  MutatorInv(FwdPtr`, \verb`Fi`, \verb`Fk`, \verb`Fl`, \verb`Ti`, \verb`Tj`, \verb`Tk`, \verb`Tl`, \verb`$toAbs`, \verb`$AbsMem`, \verb`Mem);`

\verb`  ensures  root` \verb`!=` \verb`0 ==> Pointer($toAbs`, \verb`newRoot`, \verb`old($toAbs)[root]);`

\verb`  ensures  (`$\forall^{\tt T}\!$\verb`i`$.\!$\verb` memAddr(i)` \verb`==>` \verb`$toAbs[i]` \verb`!=` \verb`$freshAbs);`

\verb`{`

\verb`  var fwd0`, \verb`fwd1`, \verb`temp;`

\verb`  assert T(root);`

\verb`  $r1 := $toAbs;`

\verb`  $r2 := MAP_NO_ABS;`

\verb`  if (root` \verb`!=` \verb`0) {`

\verb`    call newRoot := forwardFromspacePtr(root`, \verb`$freshAbs);`

\verb`  }`

\verb`  while (Tj` \verb`<` \verb`Tk)`

\verb`    invariant T(Tj) && T(root) && T(newRoot);`

\verb`    invariant GcInv(FwdPtr`,\verb`Fi`,\verb`Fk`,\verb`Fl`,\verb`Ti`,\verb`Tj`,\verb`Tk`,\verb`Tl`,\verb`$r1`,\verb`$r2`,\verb`$toAbs`,\verb`$AbsMem`,\verb`Mem);`

\verb`    invariant root` \verb`!=` \verb`0 ==> Pointer($r2`, \verb`newRoot`, \verb`old($toAbs)[root]);`

\verb`    invariant (`$\forall^{\tt T}\!$\verb`i`$.\!$\verb` memAddr(i)` \verb`==>` \verb`$toAbs[i]` \verb`!=` \verb`$freshAbs);`

\verb`  {`

\verb`    assert T(Mem[Tj,0]) && T(Mem[Tj,1]);`

\verb`    call fwd0 := forwardFromspacePtr(Mem[Tj,0]`, \verb`$freshAbs);`

\verb`    call fwd1 := forwardFromspacePtr(Mem[Tj,1]`, \verb`$freshAbs);`

\verb`    Mem[Tj,0] := fwd0;`

\verb`    Mem[Tj,1] := fwd1;`

\verb`    Tj := Tj + 1;`

\verb`  }`

\verb`  temp := Fi;`

\verb`  Fi := Ti;`

\verb`  Ti := temp;`

\verb``

\verb`  temp := Fl;`

\verb`  Fl := Tl;`

\verb`  Tl := temp;`

\verb``

\verb`  Fk := Tk;`

\verb`  Tj := Ti;`

\verb`  Tk := Ti;`

\verb``

\verb`  $toAbs := $r2;`

\verb`}`

\end{flushleft}
\caption{\small \it \textbf{Miniature Copying Collector: Garbage Collection.}}
\label{fig:exampleCopyGC}
\end{figure}

\begin{figure}
\begin{flushleft}
\verb``

\verb`procedure Alloc(root`, \verb`$abs) returns (newRoot,ptr)`

\verb`  requires MutatorInv(FwdPtr`, \verb`Fi`, \verb`Fk`, \verb`Fl`, \verb`Ti`, \verb`Tj`, \verb`Tk`, \verb`Tl`, \verb`$toAbs`, \verb`$AbsMem`, \verb`Mem);`

\verb`  requires root` \verb`!=` \verb`0 ==>`

\verb`             Pointer($toAbs`, \verb`root`, \verb`$toAbs[root]);`

\verb`  requires $abs` \verb`!=` \verb`NO_ABS;`

\verb`  requires (`$\forall^{\tt T}\!$\verb`i`$.\!$\verb` memAddr(i)` \verb`==>` \verb`$toAbs[i]` \verb`!=` \verb`$abs);`

\verb`  requires $AbsMem[$abs,0]` \verb`==` \verb`$abs;`

\verb`  requires $AbsMem[$abs,1]` \verb`==` \verb`$abs;`

\verb``

\verb`  modifies FwdPtr`, \verb`Fi`, \verb`Fk`, \verb`Fl`, \verb`Ti`, \verb`Tj`, \verb`Tk`, \verb`Tl`, \verb`$toAbs`, \verb`Mem`, \verb`$r1`, \verb`$r2;`

\verb``

\verb`  ensures  MutatorInv(FwdPtr`, \verb`Fi`, \verb`Fk`, \verb`Fl`, \verb`Ti`, \verb`Tj`, \verb`Tk`, \verb`Tl`, \verb`$toAbs`, \verb`$AbsMem`, \verb`Mem);`

\verb`  ensures  WellFormed($toAbs);`

\verb`  ensures  root` \verb`!=` \verb`0 ==>`

\verb`             Pointer($toAbs`, \verb`newRoot`, \verb`old($toAbs)[root]);`

\verb`  ensures  Pointer($toAbs`, \verb`ptr`, \verb`$abs);`

\verb``

\verb`{`

\verb`  newRoot := root;`

\verb`  assert T(root);`

\verb``

\verb`  if (Fk` \verb`>=` \verb`Fl) {`

\verb`    call newRoot := GarbageCollect(root`, \verb`$abs);`

\verb`  }`

\verb``

\verb`  while (Fk` \verb`>=` \verb`Fl) {`

\verb`    // out of memory`

\verb`  }`

\verb``

\verb`  assert T(newRoot) && T(Fk);`

\verb``

\verb`  ptr := Fk;`

\verb`  $toAbs[ptr] := $abs;`

\verb`  $r1[ptr] := $abs;`

\verb``

\verb`  Mem[ptr,0] := ptr;`

\verb`  Mem[ptr,1] := ptr;`

\verb`  FwdPtr[ptr] := 0;`

\verb``

\verb`  Fk := Fk + 1;`

\verb`}`
\end{flushleft}
\caption{\small \it \textbf{Miniature Copying Collector: Allocation.}}
\label{fig:exampleCopyAlloc}
\end{figure}

\section{A miniature copying collector in BoogiePL}\label{sec-miniature-copy}

This section applies the previous section's region-based
verification to a miniature copying collector.  Like the miniature
mark-sweep collector, the miniature copying collector is a single
BoogiePL file; it is shown in its entirety in
Figures \ref{fig:exampleCopyDef}-\ref{fig:exampleCopyAlloc}.

The copying collector is a standard two-space Cheney-queue collector
\cite{chen70}.  The heap consists of two equally sized spaces.
At any given time, one of the spaces is called {\em from-space} and the other is
called {\em to-space}.  From-space ranges from address \verb`Fi` to \verb`Fl`,
while to-space ranges from address \verb`Ti` to \verb`Tl` (where the \verb`F`
and \verb`T` stand for {\bf f}rom and {\bf t}o, and the \verb`i` and \verb`l` indicate
the {\bf i}nitial address and the {\bf l}imit of each space).

The allocator, shown in Figure \ref{fig:exampleCopyAlloc},
alloctes objects in from-space until from-space fills up.
The memory \verb`Fi`...\verb`Fk` contains allocated objects, and the
memory \verb`Fk`...\verb`Fl` contains free space, so
that allocation simply requires bumping the variable \verb`Fk`
up by one.

When from-space fills up with objects (so that \verb`Fk == Fl`), the allocator calls
the collector, shown in Figure \ref{fig:exampleCopyGC}.
The collector traverses all from-space objects reachable
from the root pointer, and copies these objects into to-space. (All objects left
in from-space are garbage, and are simply ignored by the mutator
and collector.) The collector swaps the \verb`Fi`...\verb`Fl` variables
with the \verb`Ti`...\verb`Tl` variables, so that
from-space becomes to-space and to-space becomes from-space.
The collector then returns returns control to the allocator,
which attempts allocation again.
(Note that if no garbage
existed before the collection, then no free memory will be available
after the collection; in this case, the allocator is out
of memory and has no choice but to give up.)

The \verb`forwardFromspacePointer` procedure copies a single
object, with address \verb`ptr`, from from-space to to-space.  However, before copying
the object, it checks to make sure that the object wasn't
already copied earlier.
More specifically, for each object copied to to-space, \verb`forwardFromspacePointer` sets a
{\em forwarding pointer} that points from the old from-space object
to the new to-space copy.  The variable \verb`FwdPtr` is an
array mapping each old from-space object's address to the corresponding
new to-space object address.  If \verb`FwdPtr[ptr]` is non-zero,
then the object at from-space address \verb`ptr` was already copied to the to-space address \verb`FwdPtr[ptr]`,
and \verb`forwardFromspacePointer` simply returns this to-space address.
Otherwise, \verb`forwardFromspacePointer` copies each field of
the object into to-space, sets \verb`FwdPtr[ptr]` to the to-space
address, and returns the to-space address.

When the collector copies an object to to-space, the fields of the
copied object initially point back to from-space.  The collector
later fixes up the pointers to point to to-space by
calling \verb`forwardFromspacePointer` on each field of
the to-space object.
The set of objects
not yet fixed form a contiguous work area in to-space.
The collection algorithm in Figure \ref{fig:exampleCopyGC} treats this work area as a ``scan queue'':
\verb`forwardFromspacePointer` adds newly
copied objects to the back of the queue (\verb`Tk`), and \verb`GarbageCollect` fixes objects
from the front of the queue (\verb`Tj`). When the queue is empty (\verb`Tj == Tk`), all objects
are fixed, and the collection is done.

The copying collector shares the same region-based \verb`ObjInv` from Section \ref{sec-regions}.  Other invariants differ from the mark-sweep collector, though.  For example, the copying collector has no colors, so there is no invariant to relate colors to regions.  There are invariants that relate the forwarding pointer to regions, though.
For example, each object \verb`i` in from-space satisfies this invariant,
which ensures that no object with a non-null forwarding pointer is present in \verb`$toAbs`,
and that any forwarding pointer points to a resident of \verb`$r2`:
\begin{verbatim}
   (FwdPtr[i] == 0 <==> $toAbs[i] != NO_ABS)
&& (FwdPtr[i] != 0  ==> Pointer($r2,FwdPtr[i],$r1[i]))
\end{verbatim}

The region \verb`$r2` is empty at the start of the collection.
The collector adds each object that it creates in to-space to \verb`$r2`,
while leaving \verb`$r1` unchanged.
The collector also updates \verb`$toAbs` to reflect the current
concrete location of each abstract object (either moved to to-space, or
still living in from-space);
at the end, the collector assigns \verb`$r2` to \verb`$toAbs`,
and throws out all invariants related to \verb`$r1`.

During the collection, each fixed object in to-space points from region \verb`$r2` to region \verb`$r2`:

\begin{verbatim}
    ObjInv(i,$r2,$r2,$toAbs,$AbsMem,Mem)
\end{verbatim}
Each object still in the to-space scan queue points from region \verb`$r2` back to region \verb`$r1`:

\begin{verbatim}
    ObjInv(i,$r2,$r1,$toAbs,$AbsMem,Mem)
\end{verbatim}
Meanwhile, each object in from-space points from region \verb`$r1` to region \verb`$r1`:

\begin{verbatim}
    ObjInv(i,$r1,$r1,$toAbs,$AbsMem,Mem)
\end{verbatim}
In this way, the region variables \verb`$r1` and \verb`$r2` concisely specify the state of
each object.

\section{Practical verified collectors}\label{sec-collectors}

This section applies the region-based
verification from the previous two sections to realistic copying and mark-sweep collectors,
replacing the naive recursive mark-sweep algorithm of Figures \ref{fig:exampleDefs}-\ref{fig:exampleAlloc}
with a more
efficient iterative algorithm in subsection \ref{subsec-marksweep}, then
replacing high-level language constructs with assembly language in
subsection \ref{subsec-assembly}, and then replacing the miniature 2-field, 1-root
memory model with a Bartok-compatible memory model in subsection
\ref{subsec-bartok}.  If sections \ref{sec-miniature-ms}-\ref{sec-miniature-copy} were the inspiration, this section is
the perspiration; the code for the realistic collectors is far
longer than Figures \ref{fig:exampleDefs}-\ref{fig:exampleCopyAlloc}, but not fundamentally much more
interesting.  We
present only short description and selected highlights of
the code, including a large excerpt from the realistic copying
collector in subsection \ref{subsec-copyforward}; the reader can find the full code and complete
invariants in the public release.

\subsection{Practical mark-sweep algorithm}\label{subsec-marksweep}

Our verified mark-sweep collector uses the standard 3-color
invariant.  In the beginning of the collection all objects are
white and the goal is to mark black all objects reachable from the
roots. After this marking process, the sweep process can go over
the objects to reclaim all white objects and to mark all black
objects white in preparation for the next collection. In the
beginning of the collection all objects directly reachable from
the roots are put into a list denoted the {\em mark-stack}. All
objects in this list are colored gray, meaning that they have been
reached, but their descendants have not yet been
traversed. After the roots have been scanned, the collector
proceeds by iteratively choosing a gray object $O$ from the
mark-stack, inserting $O$'s direct descendants into the mark-stack
and marking $O$ black. The black color signifies that the object
is reachable and all its direct descendants have been noticed
(i.e., put in the mark-stack). The {\em unallocated} color labels
free space.

Keeping the object color requires two bits per object. The colors
can be kept in the object header or in a separate table. Following
previous work (e.g., \cite{doma00,azat03,kerm06}) we have chosen
the latter. Bartok assumes that objects are 4-byte aligned.
Therefore, it is enough to keep two color bits per 4 bytes
(creating a space overhead of 6\%).
The two bits that correspond to the beginning of an object specify
its color. All other bit pairs are marked as unallocated. This
provides an additional benefit. When a pointer in the heap points
to a location that is marked unallocated, we know that the said
pointer is an interior pointer. Interior pointers that are
discovered during the tracing stage must be treated in a special
manner. The collector needs to find the beginning of the object in
order to discover its header and from it information on pointer
fields in the object.

The algorithm follows a very simple collection scheme. One could
choose a simpler scheme for verification, for example, by giving
up the mark-stack and searching the heap for gray objects, or
employing recursion. One could also complicate the scheme and
make it more efficient, for example, by using bit-wise sweep.
However, we attempted to find the middle way between simplicity
and efficiency, in order to enable verification while maintaining
the practicability of the collector.

\subsubsection{The allocator}
A major performance consideration is the allocator. Therefore, we
paid special attention to making the allocator efficient,
cache-friendly, scalable, and simple. We chose the local allocation
cache method that was first invented and used with the IBM JVM
allocator~\cite{borm02a} and later employed and explained in
\cite{bara05,kerm06}. This method provides efficiency by allowing
bump-pointer allocation with a mark-sweep collection. The mutator
holds a local cache in which it allocates small objects by simply
bumping a pointer. When the space in the cache is exhausted, the
mutator acquires a new local cache from the first chunk in the free
list. If that chunk is too large (larger than some threshold
maxCacheSize), then only maxCacheSize bytes of the first chunk are
taken for the local cache, and the rest is left for future cache
allocations. Allocation of large objects use the free list directly;
however, since most allocated objects in typical programs are small,
most allocation work is efficient. Furthermore, these allocations
are cache-friendly since the spatial order of allocated objects in
the memory matches the temporal order in which the program allocates
them.

Since the mutator only acquires objects or spaces of substantial
size from the free list, there is no need to keep small chunks in
it. Thus, sweep only fills the free-list with large enough spaces;
in our implementation the minimum cache size was set to 256 bytes
and objects of size 192 or up are considered large (and are thus
directly allocated from the free list).

The mark-sweep collector invariants follow the region-based approach
of Section \ref{sec-regions}, sharing the definition of \verb`Pointer` and \verb`ObjInv`
with the copying collector.  Unlike earlier sections, though,
this mark-sweep collector has a free list with non-trivial structure.
We use two ghost variables, \verb`$fs` and \verb`$fn` to represent
the size of each free list entry and the next-list-entry pointer in
each free list entry.  Any address i where \verb`$fs[i] != 0` holds
a free list entry.  Each free list entry must be at least 8 bytes:
4 bytes to store the next pointer, and 4 bytes to store the size.
The central invariant ensures, among other things,
that the space occupied by each free list entry does
not overlap with any object or any other free list entry:

\verb``

\verb`$fs[i] != 0 ==>`

\verb`    $toAbs[i] == NO_ABS`

\verb` && i + 8 <= i + $fs[i] <= memHi`

\verb` && (`$\forall^{\tt T}$\verb`j. i < j < i + $fs[i] ==>`

\verb`       $toAbs[j] == NO_ABS && $fs[j] == 0)`

\verb` && ...`

$\!\!\!\!\!\!\!\!\!\!\!$It also ensures that any non-null next-list-entry pointer points to a subsequent list entry, and that there are no other non-null next-list-entry pointers between the \verb`i` and \verb`$fn[i]`:

\verb`$fs[i] != 0 ==>`

\verb`    ($fn[i] != 0 ==> i + $fs[i] < $fn[i] <= memHi`

\verb`      && $fs[$fn[i]] != 0)`

\verb` && (`$\forall^{\tt T}$\verb`j. i < j < $fn[i]`

\verb`      && $fs[j] != 0 ==> $fn[j] == 0)`

$\!\!\!\!\!\!\!\!\!\!\!$To allocate a new local cache, the allocator
disconnects the first list element from the rest of the free list.
(For convenience in this case, the invariants allow disconnected list elements to co-exist
with the rest of the free list.)

\subsubsection{Pseudo-code}

Figure \ref{fig:mark} specifies the marking phase, written
as high-level pseudocode. The objects
reachable from the roots are marked, and then, using the
mark-stack, all reachable objects are popped from the stack,
marked black, and their children are marked gray (and pushed to
the markStack if necessary).

\begin{figure}
\CODE{
    pseudocode Mark()\\
    \> for each object Obj directly reachable from roots \\
    \> \> markStack.push(Obj); \\
    \> \> Color[Obj] := GRAY; \\
    \> Obj := markStack.pop(); \\
    \> while (Obj != null) \\
    \> \> for each object A referenced by Obj \\
    \> \> \> if (White(Color[A])) then \\
    \> \> \> \> markStack.push(A);  \\
    \> \> \> \> Color[A] := GRAY;  \\
    \> \> Color[Obj] := BLACK;\\
    \> \> Obj := markStack.pop(); \\
    \\
    pseudocode Sweep()\\
    \> Initialize free-list to null;  \\
    \> Clear any local cache currently in use; \\
    \> for (addr := heapStart; addr < heapEnd; ) \\
    \> \> color := Color[addr]; \\
    \> \> if (Black(color)) \\
    \> \> \> // Check free region prior to this object. \\
    \> \> \> regionSize := size of region since previous black object.  \\
    \> \> \> if (regionSize >= minCacheSize) \\
    \> \> \> \> Add accumulated region to free-list; \\
    \> \> \> Color[addr] := WHITE; \\
    \> \> \> addr += SizeOf(addr); \\
    \> \> elseif (White(color)) \\
    \> \> \> Color[addr] := UNALLOCATED; \\
    \> \> \> addr += SizeOf(addr); \\
    \> \> elseif (Unallocate(color)) \\
    \> \> \> addr += 4; // Step over free space. \\
  }
  \caption{\small \it \textbf{Mark-sweep pseudocode.}}
  \label{fig:mark}
\end{figure}
\begin{figure}
\CODE{
    pseudocode Allocate(size)\\
    \> If (size < largeObject) // A small object  \\
    \> \> if (size > cacheSize ) // Cache too small \\
    \> \> \> If (free-list empty) \\
    \> \> \> \>  Call GC; \\
    \> \> \> \>  if (free-list empty) \\
    \> \> \> \> \>  Out of Memory; \\
    \> \> \> Allocate cache from first free-list chunk;   \\
    \> \> // Allocate in local cache \\
    \> \> objectStart := cachePtr + cacheSize - size; \\
    \> \> cacheSize -= size; \\
    \> \> Color[objectStart] := WHITE; \\
    \> \> Initialize memory objectStart ... objectStart + size - 1 to zero. \\
    \> \> return objectStart; \\
    \> else // A large object \\
    \> \> Find first large-enough free-list chunk $C$;  \\
    \> \> If none exist then \\
    \> \> \> call GC; \\
    \> \> \> Find first large-enough free-list chunk $C$; \\
    \> \> \> If none exists then Out Of Memory; \\
    \> \> objectStart := $C$.ptr + $C$.size - size; \\
    \> \> $C$.size -= size; \\
    \> \> if ($C$.size < minCacheSize) \\
    \> \> \> remove $C$ from free-list; \\
    \> \> Color[objectStart] := WHITE; \\
    \> \> Initialize memory objectStart ... objectStart + size - 1 to zero. \\
    \> \> return objectStart; \\
  }
  \caption{\small \it \textbf{Allocation pseudocode.}}
  \label{fig:alloc}
\end{figure}

The sweep phase is also depicted in Figure \ref{fig:mark}. We keep a
very simple algorithm. Note that we do not bother maintaining
information that would allow jumping over unallocated objects
(hence the statement ``\verb`addr += 4;`'', which jumps only over one word).
There are various other optimizations possible, but we chose a
version that keeps the balance between simplicity and efficiency.

Finally, the pseudo-code of the allocator is provided in Figure
\ref{fig:alloc}. Small objects are allocated from the local
cache. A slow path occurs when the cache is exhausted or the
allocation is of a large object. In these cases the free-list must
be traversed. For a cache allocation any chunk is good, so the
first chunk is used. If the first chunk is very large, then only
part of it is assigned as the current cache.
For large objects, the list is traversed using a first-fit
allocation strategy. After the object is allocated from the chunk,
the remains of the chunk is returned to the free list only if the
created smaller chunk has size larger than minCacheSize.

\subsection{From BoogiePL to x86}\label{subsec-assembly}

So far, this paper has expressed all memory management code in
BoogiePL or in pseudocode, neither of which were designed to execute
on real computers.
We decided to write our real copying and mark-sweep collectors (and allocators) in x86 assembly language, for two reasons.
First, we didn't want a high-level language compiler in
our trusted computing base.  Second, the mutator-to-allocator
interface requires some assembly language to read the stack pointer,
so that the collectors can scan the roots on the stack.  We still
wanted to use Boogie to verify our code, so this left us with a
choice: translate annotated x86 into BoogiePL, or translate BoogiePL
into x86.  The former approach is the most common way to use
BoogiePL, but we chose the latter approach, for the following
reason.  Since the garbage collectors are written in BoogiePL, the
Boogie and Z3 tools guarantee that we really have verified the
collectors, at least in BoogiePL form, even if our BoogiePL-to-x86
translation subsequently turns the verified BoogiePL into erroneous
x86 code.  (If we had translated x86 to BoogiePL, we would have had
to ask the reader to trust that our translator didn't just
produce a trivially verifiable BoogiePL program.)

We wrote a small tool to automatically translate an x86-like subset of BoogiePL into MASM-compatible x86 code,
which we then assemble and link with Bartok-compiled benchmarks.
The tool rejects BoogiePL programs that do not conform to the
x86-like subset, such as programs that attempt to use ghost
variables at run-time.
The x86-like subset of BoogiePL (an example of which appears in Figures \ref{fig:realisticCopyA}-\ref{fig:realisticCopyB})
consists of top-level variable declarations,
non-recursive pure function declarations, and non-recursive procedure declarations.
Each procedure is either a macro that gets inline-expanded,
or a run-time procedure called with the x86 CALL instruction.
The tool enforces matching CALL and RETURN instructions;
the BoogiePL code may read the stack pointer at any time, but may not write it.
Each procedure consists of local variable declarations followed by a sequence of statements.
Since there's no recursion, local variables are statically allocated, as in early FORTRAN.
Global and local variables may be ghost variables, of any type, or physical variables, of type \verb`int`.
The tool enforces our convention that ghost variables always begin with a \verb`$` character.
The predefined global variables
\verb`eax`, \verb`ebx`, \verb`ecx`, \verb`edx`, \verb`esi`, \verb`edi`, \verb`ebp`, and \verb`esp`,
all of type \verb`int`, represent the x86 registers.
We maintain the invariant that all registers, physical variables,
and words in memory hold an integer in the range 0..$2^{32}-1$ at all times.

Each statement in a procedure is a label (used as a jump or branch target), an assignment to a ghost variable (ignored by the translation), an assignment to a register or physical variable, a procedure call, or a control statement.  Control statements are either unconditional jumps (``\verb`goto label;`'') or conditional branches:
\begin{verbatim}
if(operand1 cmp operand2) { goto label; }
\end{verbatim}
where \verb`operand1` and \verb`operand2` are registers, physical variables, or integer constants, and \verb`cmp` is a comparison operator.  Most statements are translated into single x86 instructions, but conditional branches translate into 2 x86 instructions (a compare and a branch).  A procedure call either translates into an inline expansion of the called procedure, or a single x86 CALL instruction.

Each assignment statement is either a simple move operation ``\verb`operand1 := operand2;`'', an arithmetic operation, or a memory operation.  Arithmetic operations can either statically check for 32-bit integer overflow, or check at run-time.  For example, the statement ``\verb`call eax := Sub(eax, 5);`'' statically verifies that \verb`eax - 5` does not overflow, because of the (tool-supplied) specification of \verb`Sub` (where \verb`word(e)` means that 0$\le$\verb`e`$<2^{32}$):
\begin{verbatim}
procedure Sub(x:int, y:int) returns(ret:int);
  requires word(x - y);
  ensures  ret == x - y;
\end{verbatim}
The program is not allowed to modify predefined global variables, like \verb`Mem`, directly.  To read or write memory, the program must call tool-supplied \verb`Load` and \verb`Store` procedures, which the tool translates into x86 MOV instructions.  The preconditions for \verb`Load` and \verb`Store` guarantee that the verified code does not read or write outside its allowed memory area, and that all reads and writes are to 4-byte aligned addresses.  In contrast to the two-dimensional memory \verb`Mem[objAddress,field]` presented earlier, \verb`Load` and \verb`Store` work with a
one-dimensional memory \verb`Mem[byteAddress]`.

\subsection{The Bartok memory model}\label{subsec-bartok}

Our verified garbage collectors form a critical piece of our
long-term goal: an entire verified run-time system for
Bartok-compiled code.  Because the existing Bartok run-time system
contains over 70,000 lines of code, we decided to take an
incremental approach towards creating a verified run-time system,
starting with as small a run-time system as possible, so as to
make the verification as easy as possible.  We still wanted to be
able to run real Bartok-compiled benchmarks, though, and these
benchmarks rely on many non-trivial run-time system features.  So
before attempting to verify any run-time system code, we examined
the 12 large benchmarks used in previous papers \cite{chen-bartok-tal,pizl08} to see
which features could be evicted from the run-time system.  We
found that we could remove two major features, while still
supporting 10 of the 12 benchmarks:
\begin{enumerate}[$\bullet$]
\item Only one benchmark (SpecJBB) was multithreaded, so we omitted support for multithreading from our run-time system.
\item Only one of the remaining benchmarks (mandelform) relied on GC support for unsafe code, such as pinning objects (to cast GC-managed pointers to unmanaged pointers for unsafe code) and handling callbacks from unsafe code to safe code.  Our verified GC simply halts any program that tries to use these features.
\end{enumerate}
This still left a moderately large set of features to support:
\begin{enumerate}[$\bullet$]
\item Objects have a header word, pointing to a virtual method table (vtable).  Before the header word, there is a ``pre-header'' that holds a hash code or other primitive value.
\item Non-indexed object types can have any number of fields.  Indexed object types can be strings, single-dimensional arrays, or multi-dimensional arrays, each having a different memory layout.  Array element types can be pointers, primitive values, structs without pointers, or structs with pointers.  We implemented only partial support for arrays of structs with pointers, since the 10 benchmarks did not rely on full support.
\item Pointers point to an object's header word, with one exception: root pointers may be {\em interior pointers} that point to data inside an object, ranging from the header word up to,
and including, the address of the end of the object (i.e. the address of the first word beyond the object's last field or array element).
\item An object's virtual method table has fields that the collector can read to compute the length of an object and to determine which fields of an object are pointers.  Bartok's pointer-tracking representation consists of 2 compact bit-level formats for non-indexed objects, 1 non-compact format for non-indexed objects, 1 format for strings, 2 formats for single-dimensional arrays, and 2 formats for multi-dimensional arrays.  Our collectors support all of these (except for some arrays of structs with pointers).
\item Roots may live on the stack or in static data segments.  Each static data segment has a bitmap, with one bit per static data word, indicating pointers and non-pointers in the segment.  Finding pointers on the stack is more complicated; the collector has to traverse frame pointers to find the stack frames, and it has to look up return addresses in a sorted table of return addresses to find a descriptor for each frame.  To simplify finding pointers, we set a compiler flag telling Bartok to treat all registers as caller-save registers, with no callee-save registers.
\end{enumerate}
Although the complete BoogiePL specification of the features above is
rather long and tedious, it's worth showing one example.
One of the compact pointer-tracking formats is a dense format, using one bit per field.
The specification for this says that if the tag of an object for abstract node \verb`$abs`, with vtable \verb`vt`, is \verb`DENSE_TAG`, then each field is a pointer if and only if the corresponding bit in the vtable's mask field is \verb`1`:

\verb``

\verb`tag(vt)==DENSE_TAG ==> (`$\forall^{\tt T}$\verb`j.2<=j<numFields($abs)==>`

\verb`  VFieldPtr($abs,j)==(j<30 && getBit(mask(vt),2+j)))`

\verb``

$\!\!\!\!\!\!\!\!\!\!\!$where \verb`mask` looks up a 32-bit value from the vtable (in read-only memory), and \verb`tag` and \verb`getBit` extract bits from words:
\begin{verbatim}
  fun mask(vt:int) { ro32(vt + ?VT_MASK) }
  fun tag(vt:int) { and(mask(vt), 15) }
  fun getBit(x:int,i:int) { 1 == and(shr(x, i), 1) }
\end{verbatim}
The mutator-allocator interface specification uses the uninterpreted function \verb`VFieldPtr` to state which physical values are primitive values, and which are pointer values.  The \verb`Value` function states the meaning of values in each of these two cases:
\newpage
\begin{verbatim}
  fun Value(isPtr,$r,v,$abs) {
     (isPtr && word(v) && gcAddrEx(v) && !word($abs)
                       && Pointer($r, v - 4, $abs))
  || (isPtr && word(v) && !gcAddrEx(v) && $abs == v)
  || (!isPtr && word(v) && $abs == v)
  }
\end{verbatim}
For primitive data, the data's abstract value equals its concrete value.  Pointer data may point to GC memory, under the collector's control, or they may point outside GC memory, in which case the collector treats them the same as primitive values.  The ``\verb`- 4`'' in the Pointer specification converts a pointer to a header word into a pointer to the beginning of the object (the pre-header).

Interior pointers are defined like the ordinary pointers shown above, but may have offsets larger than 4, which forces the collector to search for the beginning of the object.  The mark-sweep collector already has a table of colors, so it simply searches backwards from the interior pointer to find the first word whose color isn't \verb`unallocated`.  We also had to add an analogous bit map to the copying collector, with one bit per heap word, solely for the purpose of handling interior pointers.  (On the bright side, these bit maps did give us a chance to exercise Z3's bit vector support.)

Before we added support for Bartok's memory model, the trusted mutator-allocator specification was fairly short and readable.  After adding Bartok's memory model, the specification ballooned to hundreds of lines of bit-level details.  At this point, we started to wonder if the specification itself had bugs.  We used two techniques to test the specification.  First, we used Boogie's ``smoke'' feature, which attempts to prove false at various points in the program.  This did not turn up any bugs.  Second, we hand-translated the specification into C\# code, and then added run-time assertions to the original Bartok garbage collectors based on this C\# code.  We saw many assertion violations, which led us to 5 specification bugs, ranging from mundane (forgetting to multiply by 4 to convert a word address to a byte address) to subtle (forgetting that Bartok compresses the sorted return address tables by omitting any entry whose descriptor is identical to the previous entry).

\subsection{Example: The CopyAndForward Procedure}\label{subsec-copyforward}

As a larger example, Figures \ref{fig:realisticCopyA}-\ref{fig:realisticCopyB}
show a complete excerpt from the realistic verified copying collector:
the \verb`CopyAndForward` procedure, which copies an object
from from-space to to-space.  (This procedure corresponds to the
portion of the miniature copying collector's \verb`forwardFromspacePtr`
procedure that copies an object from from-space to to-space,
after determining that the object hasn't already been forwarded.)
In addition, the right-hand side of Figures \ref{fig:realisticCopyA}-\ref{fig:realisticCopyB}
shows the generated MASM-compatible x86 code generated by
our BoogiePL-to-x86 translation tool.

The \verb`CopyAndForward` procedure is implemented using
the control, arithmetic, and memory constructs described
in subsection \ref{subsec-assembly}: \verb`if`, \verb`goto`,
\verb`call`, \verb`AddChecked`, \verb`Sub`, etc.  There's one slight embellishment
in the implementation: the x86-like subset of BoogiePL distinguishes
between the read-only memory that describes Bartok-generated
GC tables, the read-write stack memory that the mutator controls,
and the read-write heap memory that the garbage collector controls.
The variable \verb`$GcMem` represents the last of these, and
the garbage collector uses \verb`GcLoad` and \verb`GcStore`
operations to read and write \verb`$GcMem`.  The translator
turns \verb`GcLoad` and \verb`GcStore` into x86 \verb`mov`
instructions, as seen on the right-hand side of Figures
\ref{fig:realisticCopyA}-\ref{fig:realisticCopyB}.

The \verb`CopyAndForward` procedure relies on several helper
procedures, also written in BoogiePL and verified using
Boogie/Z3 (and all available in the public source release).  The
\verb`GetSize` procedure accepts a pointer in register \verb`ecx`
to an object with vtable \verb`edx`, and computes the size
of the object.  (This is complicated in general, because
the object may be a non-index type, a string, or a single- or
multi-dimensional array.)  The inline procedure \verb`copyWord`
copies a single field, with field index \verb`edi`,
from from-space object \verb`ecx` to to-space object \verb`esi`.
(We split this into a separate procedure, because the
verification time of the separate procedures was lower
than the verification time of a single, combined procedure.)
Finally, the inline procedure \verb`bb4SetBit` sets a single
bit, at position \verb`esi`, in a bit-vector at address \verb`edi`.
(The bit vector consists of an array of 4-byte words,
each containing 32 bits of the bit vector.)  Note that
the translator inlines the code
from \verb`copyWord` and \verb`bb4SetBit` directly into
the code for \verb`CopyAndForward`, as seen on the right-hand side of Figures
\ref{fig:realisticCopyA}-\ref{fig:realisticCopyB}.

The miniature collectors used a trigger \verb`T` in quantifiers
to guide quantifier instantiation.  To reduce unnecessary
quantifier instantiation,
the realistic collector implementation uses seperate
triggers for separate purposes: \verb`TV` is used for
general-purpose values, including pointers, while \verb`TO`
is used for field indices.

The preconditions to \verb`CopyAndForward` specify the following:
\begin{enumerate}[$\bullet$]
\item The \verb`ecx` register contains a pointer \verb`$ptr`,
which is a valid pointer to a from-space object.
\item The copying collector's overall invariant \verb`CopyGcInv`
on GC memory holds.  (This invariant is like the \verb`GcInv`
invariant in the miniature copying collector, although it deals
with more complexities than the miniature collector.  For
example, the object at address \verb`Tj`, the head of the scan queue,
may be in the middle of being scanned as \verb`CopyAndForward`
runs.  The \verb`CopyGcInv` keeps track of both the beginning
of this head object, \verb`Tj`, and the end of the head object, \verb`$_tj`.)
\item The from-space object has not been forwarded (\verb`!IsFwdPtr(...)`).
Note that unlike the miniature copying collector,
the real collector stores the forwarding pointer in the
header field of a from-space object after the from-space
object is copied, overwriting the vtable (virtual method table)
pointer in the header.
(The collector can distinguish a vtable pointer from a
forwarding pointer, because vtables do not live in to-space.)
Also note that the header field follows the pre-header field,
so it lives at address \verb`$ptr + 4` rather than \verb`$ptr`.
\item The object has been reached.  (This is used to prove
that all copied objects were reached during the collection,
so that non-reached objects are actually collected.)
\end{enumerate}

To copy an object, \verb`CopyAndForward` first loads the vtable
from the object's header and calls \verb`GetSize` to get the size
of the object, which it places in \verb`ebp`.  It then
reserves space for the copied object in to-space, by
adding \verb`ebp` bytes to the to-space scan queue tail \verb`Tk`
and checking that this addition causes neither a 32-bit integer
overflow nor an overflow past the to-space limit \verb`Tl`.
(With additional effort, we could probably prove that enough
space will always be available in to-space, but the run-time
cost of checking for space is small.)

After reserving memory in to-space, \verb`CopyAndForward`
enters a loop that copies each field \verb`edi` of the object.
The ``\verb`assert`'' statements after the \verb`loop` label
specify the loop invariants.  (For conciseness, Figure
\ref{fig:realisticCopyB} omits most of the loop invariants,
which are similar to \verb`CopyAndForward`'s postconditions, but
longer.)

After copying the object, \verb`CopyAndForward` overwrites
the old from-space object's vtable with a forwarding pointer
to the new to-space object.  (Note that the x86 load-effective-address
instruction \verb`lea` simply computes an address.)
Next, \verb`CopyAndForward` sets a bit in the GC bit vector
to indicate the start of an object, so that the collector
can later find the start of the object from an interior pointer.
Finally, \verb`CopyAndForward` updates the ghost variables
\verb`$r2` and \verb`$toAbs` and returns.

At the end of \verb`CopyAndForward`, the postconditions guarantee that:
\begin{enumerate}[$\bullet$]
\item The overall invariant \verb`CopyGcInv` still holds
\item The new \verb`$r2` is a valid extension of the old \verb`$r2`
\item The return value, \verb`eax`, is a valid pointer to a to-space
object.  Furthermore, the fields of the object point back to from-space.
(In region terminology, the object points from the to-space region \verb`$r2`
to the from-space region \verb`$r1`.)
\item The to-space scan queue grows at the tail (\verb`Tk`), but
remains the same at the head (\verb`Tj`).
\item The object at the head of the to-space queue (\verb`Tj`) is unmodified.
\item The to-space object pointed to by \verb`eax` is actually
located in the to-space memory area \verb`Ti`...\verb`Tl`.
\end{enumerate}

\begin{table}[t]
\begin{tabular}{|l|c|c|c|}
\hline
               & BoogiePL code    & x86                   & Time to       \\
               & (non-comment,    & instructions          & verify        \\
               & non-blank lines) & (before inlining)     & (seconds)     \\
\hline
Trusted defs & 546 & & \\
\hline
Shared code & 779 & 177 & 12 \\
\hline
Copying & 2398 & 802 & 115 \\
\hline
Mark-sweep & 3038 & 865 & 70 \\
\hline
\end{tabular}
\caption{Verification times for practical collectors}
\label{table:vertime}
\end{table}

\begin{figure}
\begin{flushleft}
$\!\!\!\!\!\!\!\!$\verb`procedure CopyAndForward($ptr`, \verb`$_tj)`

$\!\!\!\!\!\!\!\!$\verb`  requires ecx` \verb`==` \verb`$ptr;`

$\!\!\!\!\!\!\!\!$\verb`  requires CopyGcInv(...);`

$\!\!\!\!\!\!\!\!$\verb`  requires Pointer($r1`, \verb`$ptr`, \verb`$r1[$ptr]) && TV($ptr);`

$\!\!\!\!\!\!\!\!$\verb`  requires !IsFwdPtr($GcMem[$ptr + 4]);`

$\!\!\!\!\!\!\!\!$\verb`  requires $_tj` \verb`<=` \verb`Tk;`

$\!\!\!\!\!\!\!\!$\verb`  requires reached($toAbs[$ptr]`, \verb`$Time);`

$\!\!\!\!\!\!\!\!$\verb`  modifies $r2`, \verb`$GcMem`, \verb`$toAbs`, \verb`Tk`, \verb`$gcSlice;`

$\!\!\!\!\!\!\!\!$\verb`  modifies eax`, \verb`ebx`, \verb`ecx`, \verb`edx`, \verb`esi`, \verb`edi`, \verb`ebp`, \verb`esp;`

$\!\!\!\!\!\!\!\!$\verb`  ensures  CopyGcInv(...);`

$\!\!\!\!\!\!\!\!$\verb`  ensures  RExtend(old($r2)`, \verb`$r2);`

$\!\!\!\!\!\!\!\!$\verb`  ensures  Pointer($r2`, \verb`eax`, \verb`$r1[$ptr]);`

$\!\!\!\!\!\!\!\!$\verb`  ensures  Tj` \verb`==` \verb`old(Tj);`

$\!\!\!\!\!\!\!\!$\verb`  ensures  Tk` \verb`>=` \verb`old(Tk);`

$\!\!\!\!\!\!\!\!$\verb`  ensures  old($toAbs)[Tj]` \verb`!=` \verb`NO_ABS ==>`

$\!\!\!\!\!\!\!\!$\verb`             $toAbs[Tj]` \verb`!=` \verb`NO_ABS && $toAbs[Tj]` \verb`==` \verb`old($toAbs)[Tj];`

$\!\!\!\!\!\!\!\!$\verb`  ensures  (forall j::{TO(j)} TO(j) ==>`

$\!\!\!\!\!\!\!\!$\verb`              0` \verb`<=` \verb`j && Tj + 4 * j` \verb`<` \verb`$_tj ==>`

$\!\!\!\!\!\!\!\!$\verb`                $GcMem[Tj + 4 * j]` \verb`==` \verb`old($GcMem)[Tj + 4 * j] && ...;`

$\!\!\!\!\!\!\!\!$\verb`  ensures  Ti` \verb`<=` \verb`eax && eax` \verb`<` \verb`Tk && gcAddrEx(eax + 4);`

$\!\!\!\!\!\!\!\!$\verb`{`

$\!\!\!\!\!\!\!\!$\verb`  var tmp;`

$\!\!\!\!\!\!\!\!$\verb``

$\!\!\!\!\!\!\!\!$\verb`  call edx := GcLoad(ecx + 4);             ` $\Longrightarrow$ \tt \bf  mov edx,dword ptr [ecx+4] \verb``

$\!\!\!\!\!\!\!\!$\verb`  esp := esp - 4; call GetSize($ptr`, \verb`...); `\, $\Longrightarrow$ \tt \bf  call \_?GetSize \verb``

$\!\!\!\!\!\!\!\!$\verb`  ebp := eax;                              ` $\Longrightarrow$ \tt \bf  mov ebp,eax \verb``

$\!\!\!\!\!\!\!\!$\verb`  assert TO(numFields($r1[$ptr]));`

$\!\!\!\!\!\!\!\!$\verb``

$\!\!\!\!\!\!\!\!$\verb`  esi := Tk;                               ` $\Longrightarrow$ \tt \bf  mov esi,dword ptr \_\$\$Tk \verb``

$\!\!\!\!\!\!\!\!$\verb`  call Tk := AddChecked(Tk`, \verb`ebp);          `\,\, $\Longrightarrow$ \tt \bf  add dword ptr \_\$\$Tk,ebp \verb``

$\!\!\!\!\!\!\!\!$\verb`                                               ` \tt \bf  jc overflowed \verb``

$\!\!\!\!\!\!\!\!$\verb`  assert TV(esi);`

$\!\!\!\!\!\!\!\!$\verb``

$\!\!\!\!\!\!\!\!$\verb`  eax := Tl;                               ` $\Longrightarrow$ \tt \bf  mov eax,dword ptr \_\$\$Tl \verb``

$\!\!\!\!\!\!\!\!$\verb`  if (Tk` \verb`<=` \verb`eax) { goto skip1; }           `\,\, $\Longrightarrow$ \tt \bf  cmp dword ptr \_\$\$Tk,eax \verb``

$\!\!\!\!\!\!\!\!$\verb`                                               ` \tt \bf  jbe CopyAndForward\$skip1 \verb``

$\!\!\!\!\!\!\!\!$\verb`    // out of memory`

$\!\!\!\!\!\!\!\!$\verb`    call DebugBreak();                     ` $\Longrightarrow$ \tt \bf  int 3 \verb``

$\!\!\!\!\!\!\!\!$\verb`  skip1:                                   ` $\Longrightarrow$ \tt \bf  CopyAndForward\$skip1: \verb``

$\!\!\!\!\!\!\!\!$\verb``

$\!\!\!\!\!\!\!\!$\verb`  edi := 0;                                ` $\Longrightarrow$ \tt \bf  mov edi,0 \verb``

$\!\!\!\!\!\!\!\!$\verb`  edx := 0;                                ` $\Longrightarrow$ \tt \bf  mov edx,0 \verb``
\end{flushleft}
\caption{\small \it \textbf{Realistic Copying Collector: forwarding, part 1/2.}}
\label{fig:realisticCopyA}
\end{figure}
\begin{figure}
\begin{flushleft}
$\!\!\!\!\!\!\!\!$\verb`  loop:                                    ` $\Longrightarrow$ \tt \bf  CopyAndForward\$loop: \verb``

$\!\!\!\!\!\!\!\!$\verb`      // loop invariants:`

$\!\!\!\!\!\!\!\!$\verb`      assert 4 * edi` \verb`==` \verb`edx;`

$\!\!\!\!\!\!\!\!$\verb`      assert TO(edi) && 0` \verb`<=` \verb`edi && edi` \verb`<=` \verb`numFields($r1[$ptr]);`

$\!\!\!\!\!\!\!\!$\verb`      assert CopyGcInv(...);`

$\!\!\!\!\!\!\!\!$\verb`      assert RExtend(old($r2)`, \verb`$r2);`

$\!\!\!\!\!\!\!\!$\verb`      ...`

$\!\!\!\!\!\!\!\!$\verb`    if (edx` \verb`>=` \verb`ebp) { goto loopEnd; }      ` $\Longrightarrow$ \tt \bf  cmp edx,ebp \verb``

$\!\!\!\!\!\!\!\!$\verb`                                              `\,  \tt \bf  jae CopyAndForward\$loopEnd \verb``

$\!\!\!\!\!\!\!\!$\verb`    // copy one field:`

$\!\!\!\!\!\!\!\!$\verb`    call copyWord($ptr`,\verb`$_tj`,\verb`esi`,\verb`edi`,\verb`ebp);   ` $\Longrightarrow$ \tt \bf  mov eax,dword ptr [ecx+4*edi] \verb``

$\!\!\!\!\!\!\!\!$\verb`                                              `\, \tt \bf  mov dword ptr [esi+4*edi],eax \verb``

$\!\!\!\!\!\!\!\!$\verb`    call edi := Add(edi`, \verb`1);               ` $\Longrightarrow$ \tt \bf  add edi,1 \verb``

$\!\!\!\!\!\!\!\!$\verb`    call edx := Add(edx`, \verb`4);               ` $\Longrightarrow$ \tt \bf  add edx,4 \verb``

$\!\!\!\!\!\!\!\!$\verb`    goto loop;                            ` $\Longrightarrow$ \tt \bf  jmp CopyAndForward\$loop \verb``

$\!\!\!\!\!\!\!\!$\verb`  loopEnd:                                ` $\Longrightarrow$ \tt \bf  CopyAndForward\$loopEnd: \verb``

$\!\!\!\!\!\!\!\!$\verb``

$\!\!\!\!\!\!\!\!$\verb`  // set forwarding pointer:`

$\!\!\!\!\!\!\!\!$\verb`  call eax := Lea(esi + 4);               ` $\Longrightarrow$ \tt \bf  lea eax,dword ptr [esi+4] \verb``

$\!\!\!\!\!\!\!\!$\verb`  call GcStore(ecx + 4`, \verb`eax);             `\, $\Longrightarrow$ \tt \bf  mov dword ptr [ecx+4],eax \verb``

$\!\!\!\!\!\!\!\!$\verb`  eax := esi;                             ` $\Longrightarrow$ \tt \bf  mov eax,esi \verb``

$\!\!\!\!\!\!\!\!$\verb`  // set bit in table:`

$\!\!\!\!\!\!\!\!$\verb`  call esi := Sub(esi`, \verb`Ti);                ` $\Longrightarrow$ \tt \bf  sub esi,dword ptr \_\$\$Ti \verb``

$\!\!\!\!\!\!\!\!$\verb`  edi := BT;                              ` $\Longrightarrow$ \tt \bf  mov edi,dword ptr \_\$\$BT \verb``

$\!\!\!\!\!\!\!\!$\verb`  call bb4SetBit(...);                    ` $\Longrightarrow$ \tt \bf  mov ecx,esi \verb``

$\!\!\!\!\!\!\!\!$\verb`                                              ` \tt \bf  shr esi,7 \verb``

$\!\!\!\!\!\!\!\!$\verb`                                              ` \tt \bf  shr ecx,2 \verb``

$\!\!\!\!\!\!\!\!$\verb`                                              ` \tt \bf  add esi,esi \verb``

$\!\!\!\!\!\!\!\!$\verb`                                              ` \tt \bf  add esi,esi \verb``

$\!\!\!\!\!\!\!\!$\verb`                                              ` \tt \bf  add esi,edi \verb``

$\!\!\!\!\!\!\!\!$\verb`                                              ` \tt \bf  and ecx,31 \verb``

$\!\!\!\!\!\!\!\!$\verb`                                              ` \tt \bf  mov edi,1 \verb``

$\!\!\!\!\!\!\!\!$\verb`                                              ` \tt \bf  shl edi,cl \verb``

$\!\!\!\!\!\!\!\!$\verb`                                              ` \tt \bf  mov ecx,edi \verb``

$\!\!\!\!\!\!\!\!$\verb`                                              ` \tt \bf  mov edi,dword ptr [esi] \verb``

$\!\!\!\!\!\!\!\!$\verb`                                              ` \tt \bf  or edi,ecx \verb``

$\!\!\!\!\!\!\!\!$\verb`                                              ` \tt \bf  mov dword ptr [esi],edi \verb``

$\!\!\!\!\!\!\!\!$\verb`  $r2[eax] := $r1[$ptr];`

$\!\!\!\!\!\!\!\!$\verb`  $toAbs[eax] := $toAbs[$ptr];`

$\!\!\!\!\!\!\!\!$\verb`  $toAbs[$ptr] := NO_ABS;`

$\!\!\!\!\!\!\!\!$\verb`  assert TO(1);`

$\!\!\!\!\!\!\!\!$\verb`  assert TV(eax - Ti);`

$\!\!\!\!\!\!\!\!$\verb``

$\!\!\!\!\!\!\!\!$\verb`  esp := esp + 4; return;                  ` $\Longrightarrow$ \tt \bf  ret \verb``

$\!\!\!\!\!\!\!\!$\verb`}`
\end{flushleft}
\caption{\small \it \textbf{Realistic Copying Collector: forwarding, part 2/2.}}
\label{fig:realisticCopyB}
\end{figure}

\section{Performance}\label{sec-measurements}

\begin{figure}[p]
 \begin{center}
  \subfigure[Othello]{\label{fig:performance-othello}\includegraphics[viewport=49 162 743 448,clip,width=0.495\linewidth]{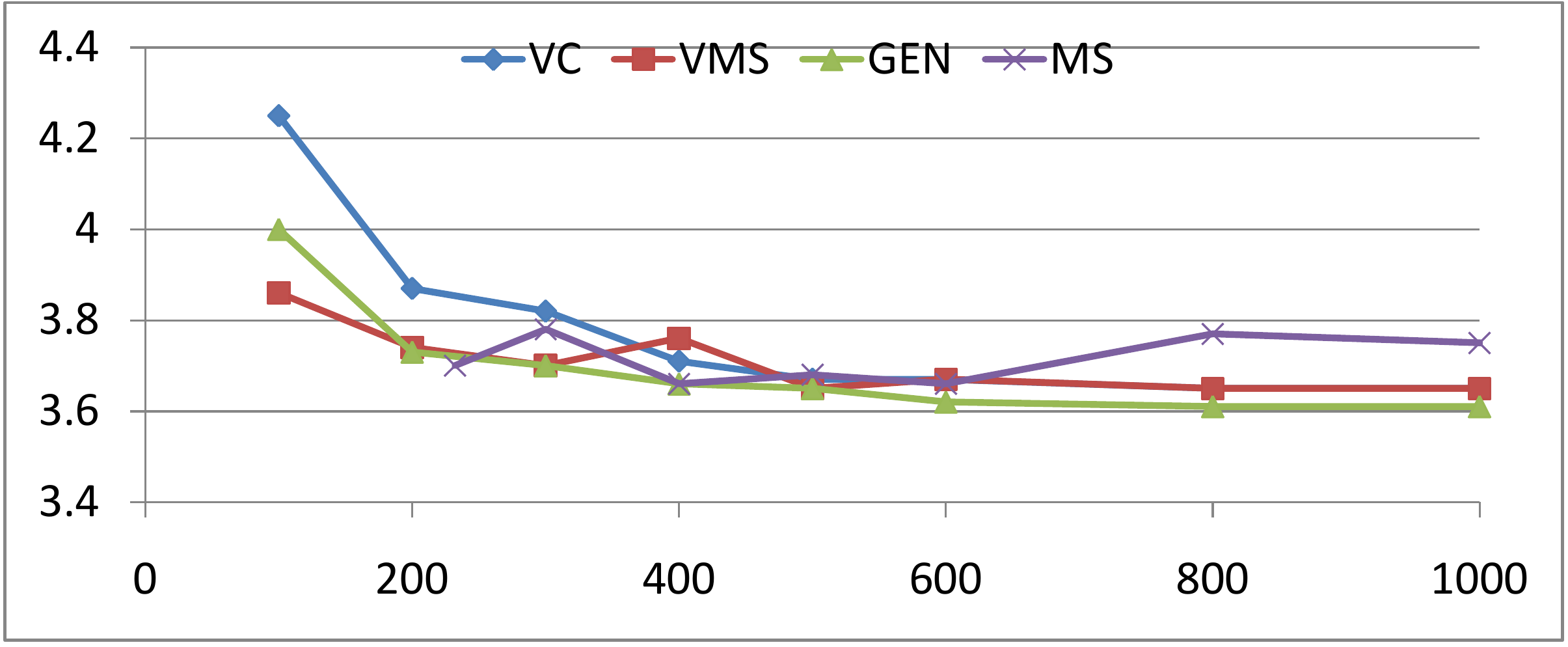}}
  \subfigure[Ahc]{\label{fig:performance-ahc}\includegraphics[viewport=49 162 743 448,clip,width=0.495\linewidth]{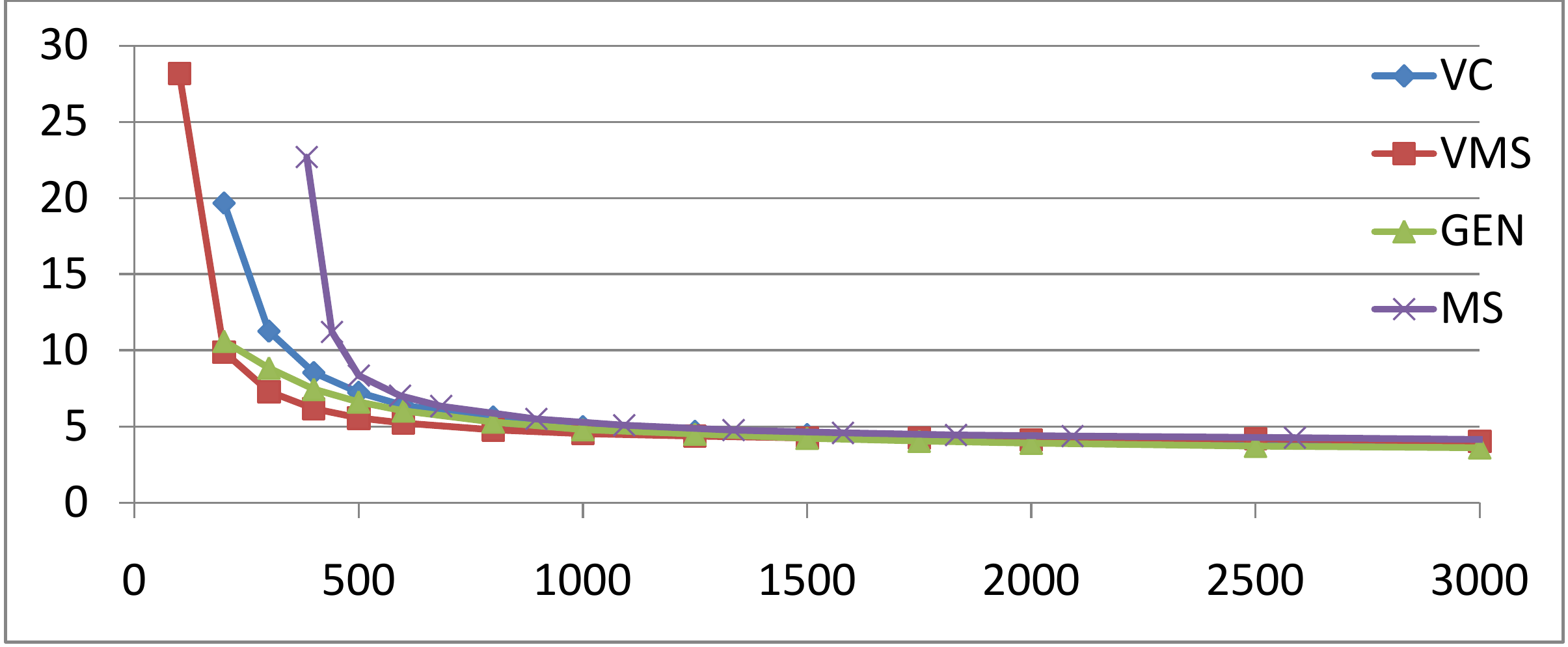}}
  \subfigure[Go]{\label{fig:performance-go}\includegraphics[viewport=49 162 743 448,clip,width=0.495\linewidth]{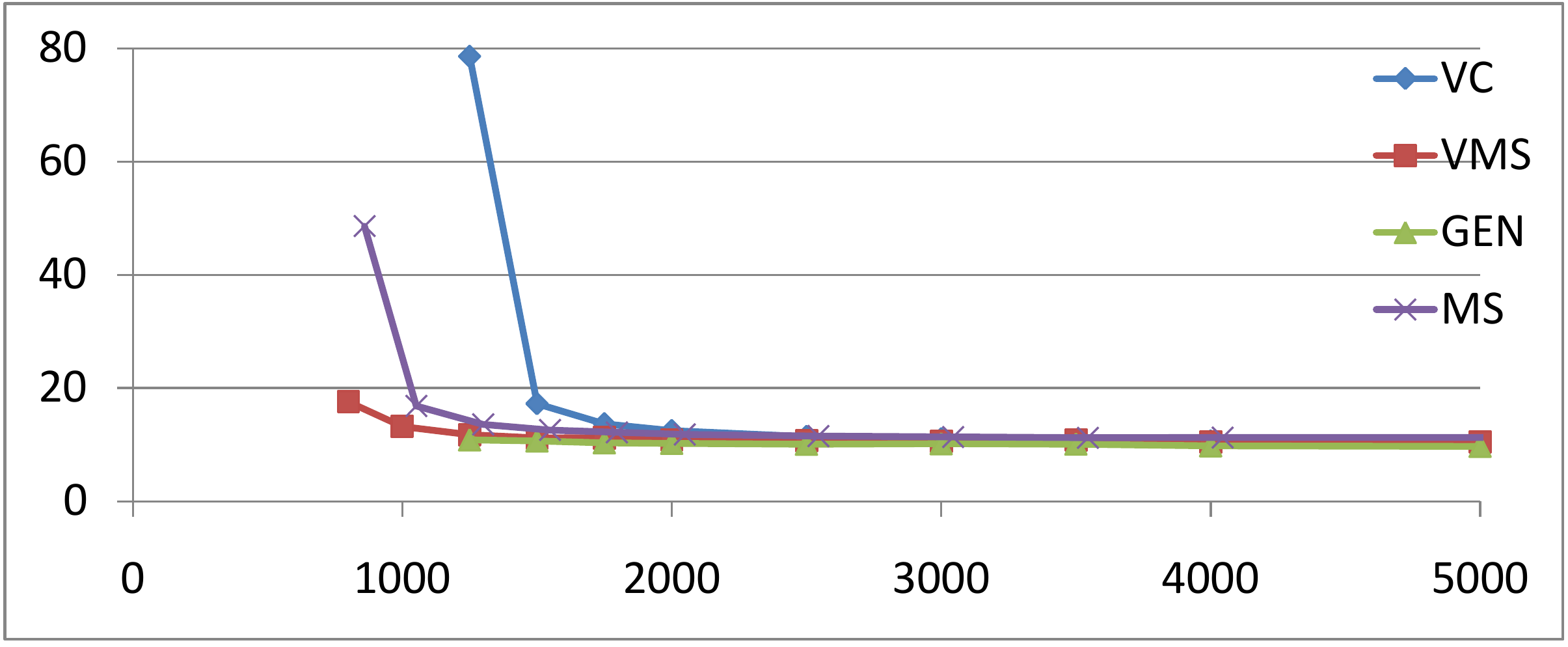}}
  \subfigure[Xlisp]{\label{fig:performance-xlisp}\includegraphics[viewport=49 162 743 448,clip,width=0.495\linewidth]{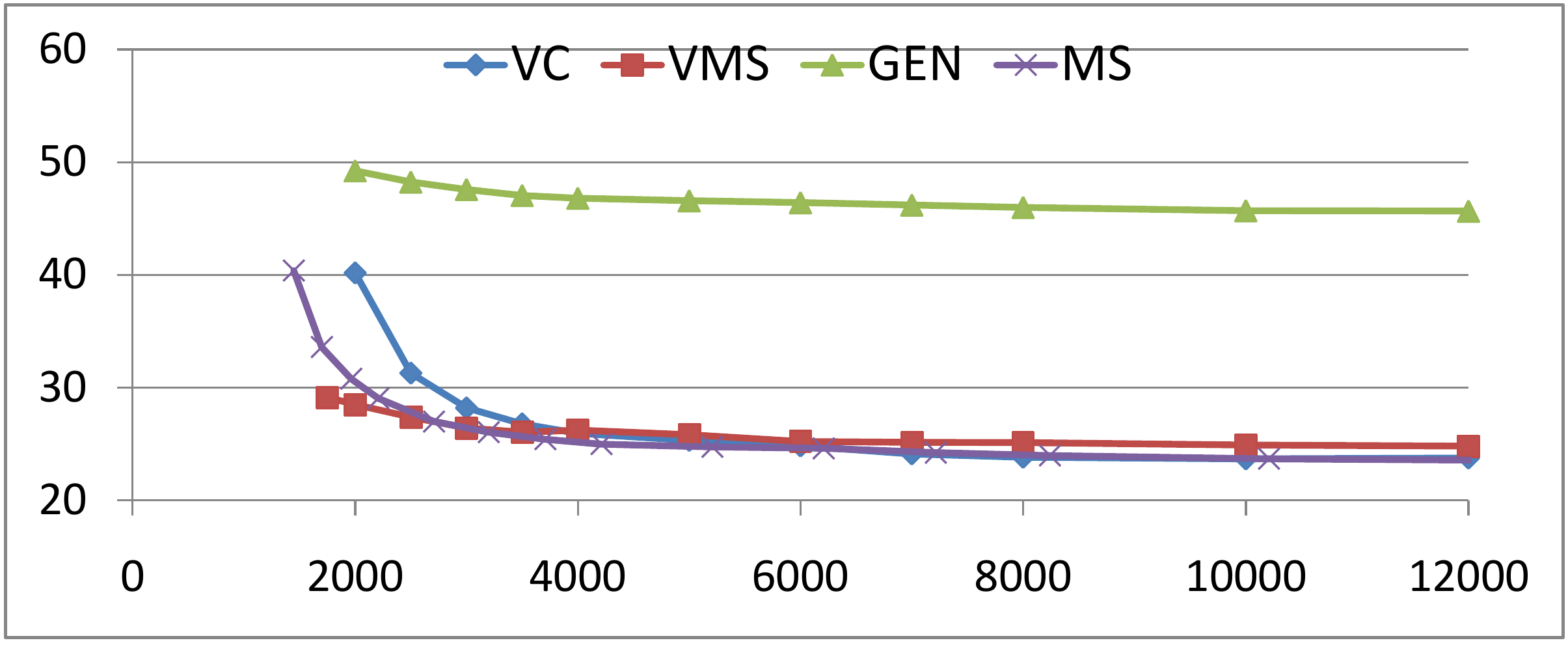}}
  \subfigure[Crafty]{\label{fig:performance-crafty}\includegraphics[viewport=49 162 743 448,clip,width=0.495\linewidth]{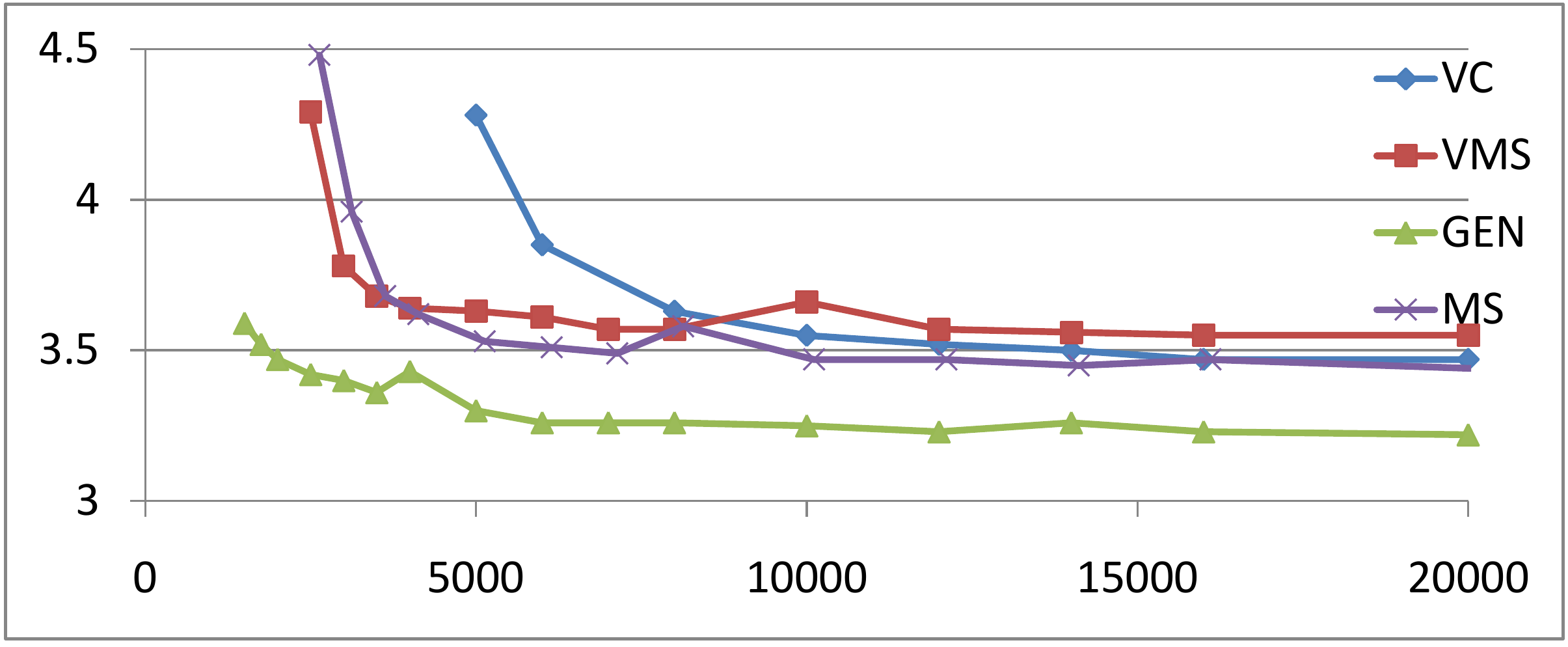}}
  \subfigure[Zinger]{\label{fig:performance-zinger}\includegraphics[viewport=49 162 743 448,clip,width=0.495\linewidth]{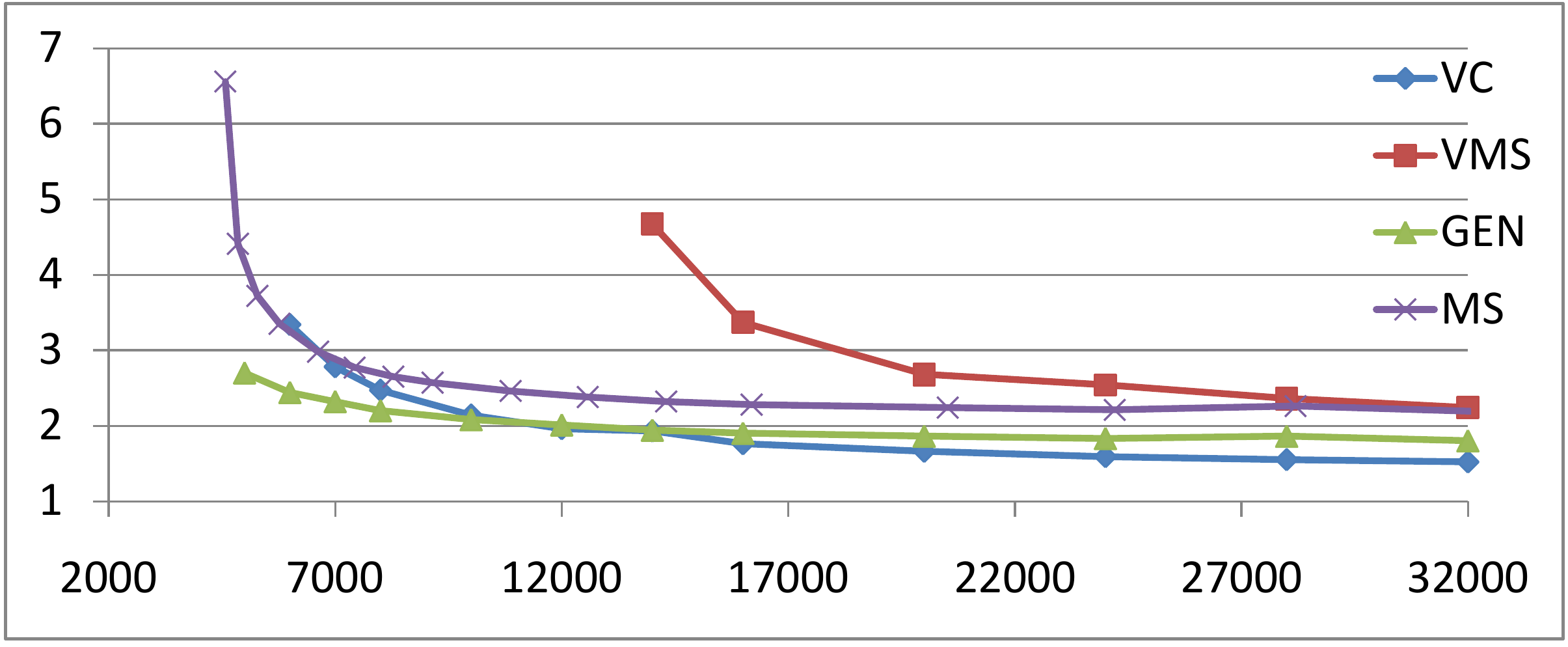}}
  \subfigure[Sat]{\label{fig:performance-sat}\includegraphics[viewport=49 162 743 448,clip,width=0.495\linewidth]{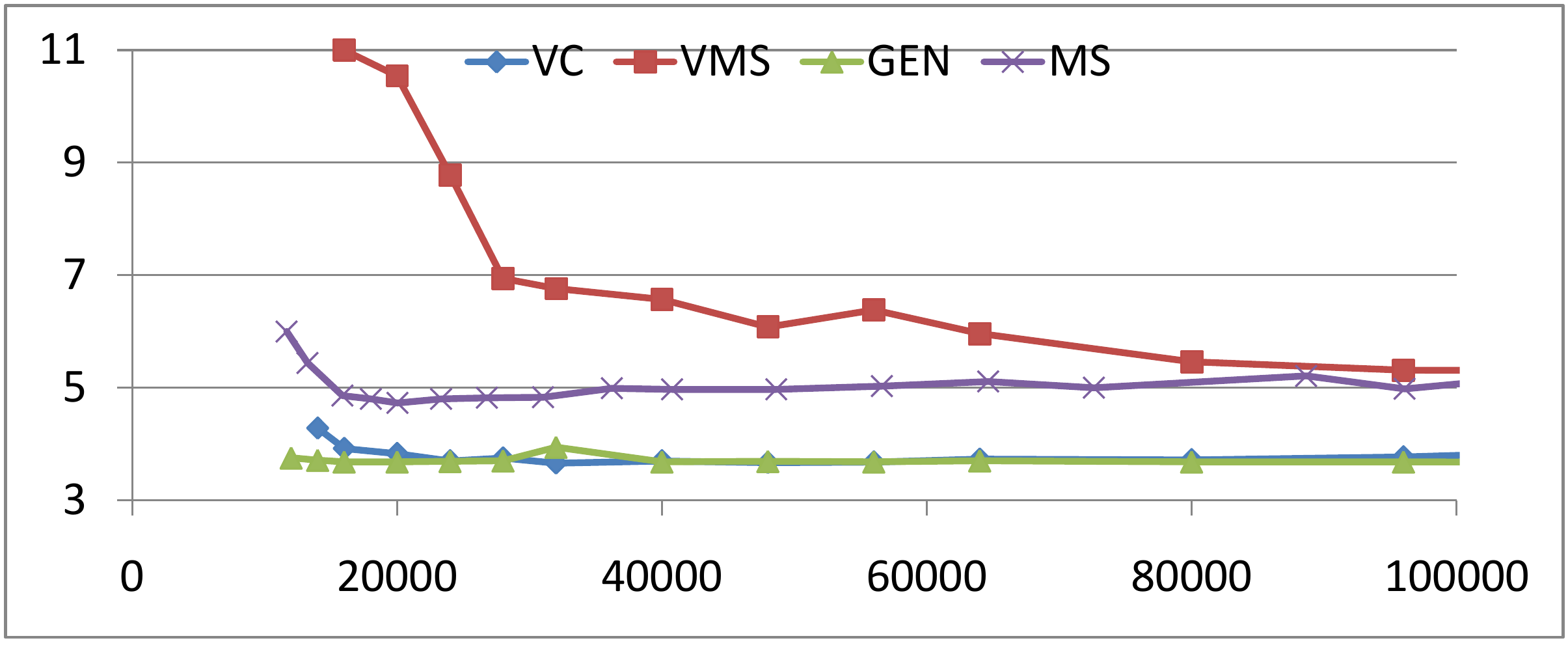}}
  \subfigure[Asmlc]{\label{fig:performance-asmlc}\includegraphics[viewport=49 162 743 448,clip,width=0.495\linewidth]{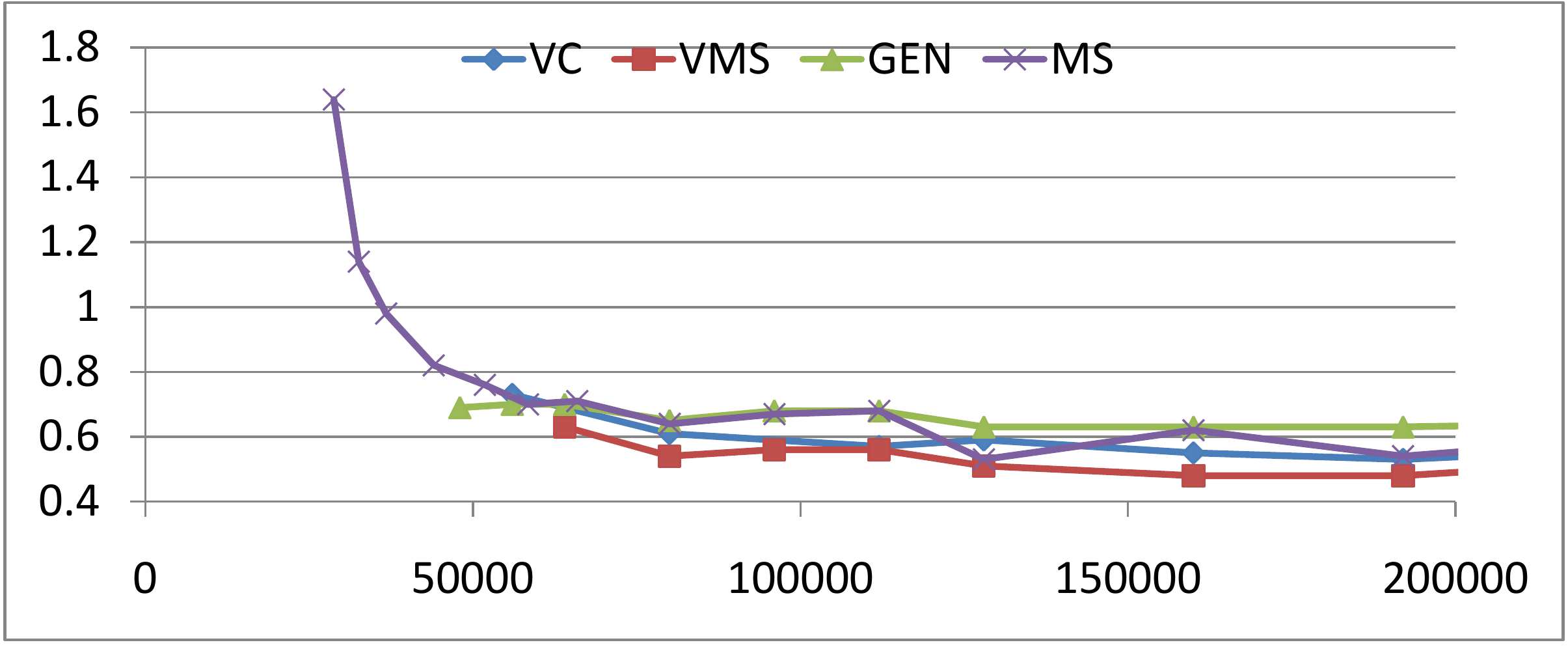}}
  \subfigure[Lcsc]{\label{fig:performance-lcsc}\includegraphics[viewport=49 162 743 448,clip,width=0.495\linewidth]{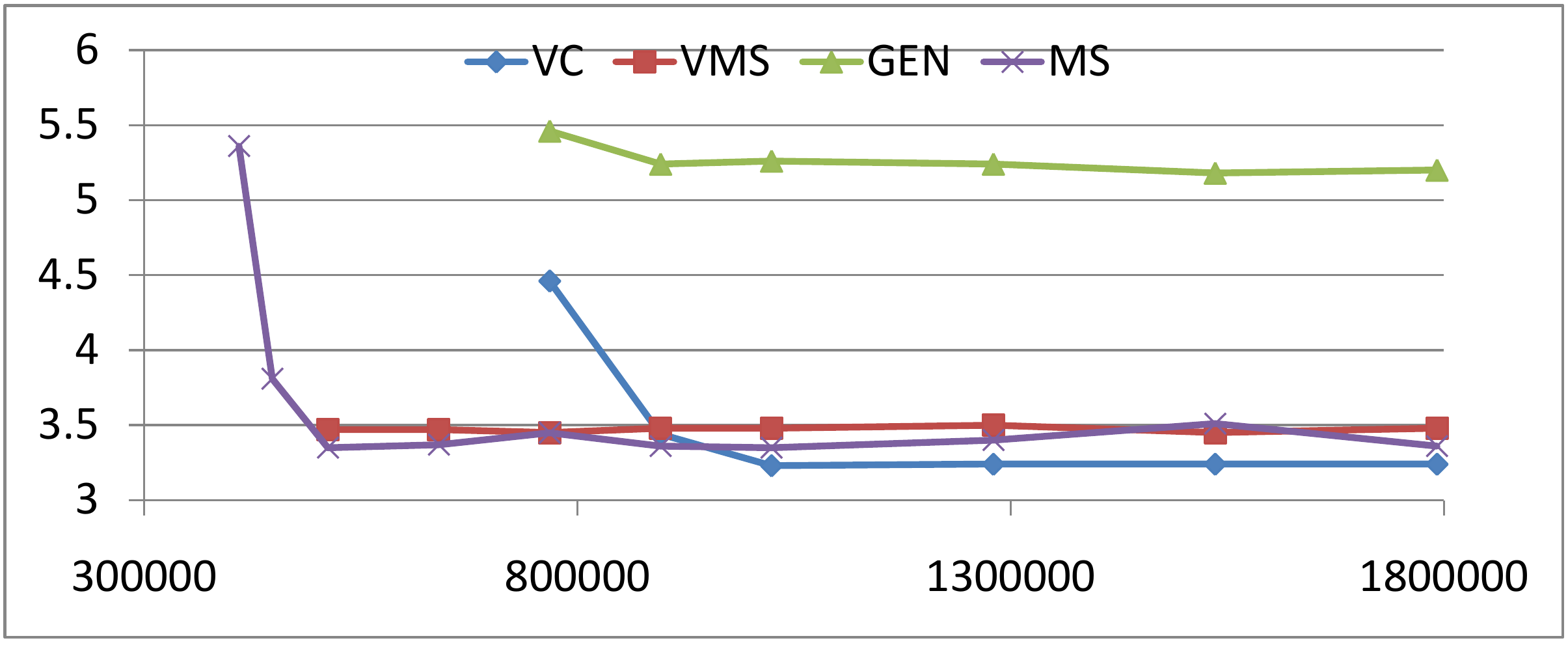}}
  \subfigure[Bartok]{\label{fig:performance-bartok}\includegraphics[viewport=49 162 743 448,clip,width=0.495\linewidth]{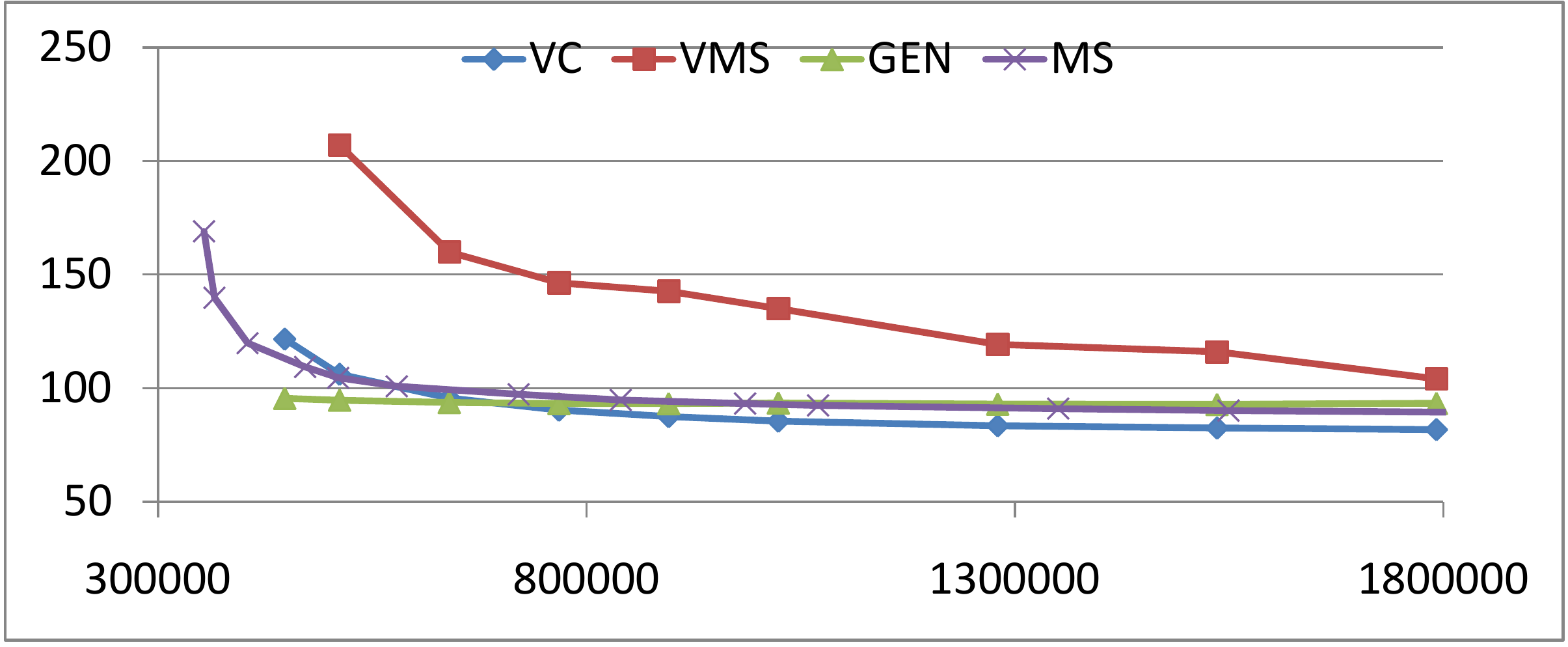}}
 \end{center}
 \caption{Performance Comparisons: overall running time (in seconds) vs. heap size (KB)}
 \label{fig:performance}
\end{figure}

This section presents performance results, measured on a single
core of a 4-core, 2.4GHz Intel Core2 with 4GB of RAM, 4MB of L2
cache, and a 64KB L1 cache.

Verifying the copying collector, mark-sweep collector, and
the code shared between the collectors took 115
seconds, 70 seconds, and 12 seconds, respectively
(see Table \ref{table:vertime}).  This fast verification
reflects our choice of a simple trigger \verb`T(i)`.  The copying
collector and mark-sweep collector contained 802 x86 instructions
(before inlining) and 865 x86 instructions (before inlining), plus
177 x86 instructions (before inlining) shared between the collectors.
The BoogiePL files for the copying and mark-sweep
collectors contained 2398 non-comment, non-blank lines and 3038
non-comment, non-blank lines, plus 779 non-comment, non-blank
lines of BoogiePL code shared between the collectors.
Thus, there are about 2-3 lines of annotation per x86 instruction.
These annotations require a non-trivial amount of human effort to write,
but the effort is not too much greater than the effort spent in ordinary development and testing.
The trusted
definitions, including x86 instruction specifications and the
Bartok interface specification, occupied 546 non-blank,
non-comment lines.

Figure \ref{fig:performance} shows the performance of
the 10 benchmarks cited in Section \ref{sec-collectors} as a
function of heap size, both for our verified memory managers and
for Bartok's native run-time system. We denote the verified
copying collector by {\sc vc}, the verified mark-sweep collector
by {\sc vms}, the generational copying Bartok collector by {\sc
gen}, and the Bartok standard mark-sweep collector by {\sc ms}.
These results demonstrate that (a) our collectors work on real
benchmarks, and (b) the space and time consumption is in the same
ballpark as Bartok's native run-time system.  We emphasize the
``ballpark'' nature of the comparison between the verified
collectors and the native Bartok collectors, because this
comparison is highly unfair to the native collectors, which
support more features than the verified collectors.  In
particular, Bartok's native run-time system supports
multithreading, which adds synchronization overhead to the
mutator and memory manager.

Bartok's native collectors were not designed to be used with a
fixed heap size; they expect to grow the heap as needed.  To get a
time vs. space plot for the Bartok collectors, we varied the
triggering mechanism used for heap growth, and then measured the
actual heap space used.  For the generational collector, we set
the nursery size to 4MB or 1/4 of the maximum heap size, whichever
was smaller.

Several benchmarks created fragmentation that made it difficult for
the verified mark-sweep collector to find space for very large
objects.  The standard Bartok mark-sweep collector simply grows the
heap when the current heap lacks space for a very large object; we
configured the verified mark-sweep collector to set aside part of
the heap as a wilderness area, used as a last resort for very large
object allocation.  While this wilderness area enabled these
benchmarks to keep running under heavy fragmentation, the
performance still suffered, as seen in figures~\ref{fig:performance-zinger}, \ref{fig:performance-sat}, and \ref{fig:performance-bartok}.
For other benchmarks, though, the
verified mark-sweep collector performed well across a large spectrum
of heap sizes.  The verified copying collector, as expected,
required a larger minimum heap size, but performed asymptotically
well as the heap size increased.

\section{Conclusion}\label{sec-conslusion}
We have presented two simple collectors with the minimal set
of properties required to make them reasonably efficient in a
practical setting.  We have mechanically verified that these
collectors maintain a heap representation that is faithful
to a mutator-defined abstract heap, and have run the
collector on large, off-the-shelf C\# benchmarks.

Given the large size of the mutator-allocator specification, we were very curious to see whether our collectors would run correctly the first time.  Alas, running the verified copying collector revealed two specification bugs that we hadn't caught before: \verb`Initialize`'s postcondition forgot to ensure that the \verb`ebp` register was saved, and the allocation postcondition specified a return value that was off by 4 bytes (a header/pre-header confusion).  Thus, the copying collector ran correctly the {\em third} time we tried it, which is still no small achievement for a garbage collector hand-coded in assembly language.  Furthermore, we were then able to verify the mark-sweep collector against the debugged specification, so that the mark-sweep collector ran correctly the first time we tried it.  In addition, having a clear and well-tested specification is useful for TAL/PCC verifiers: based on the specification, we found a bug in our TAL verifier \cite{chen-bartok-tal}, which didn't check that the sparse pointer tracking formats mention no field more than once; this bug can allow TAL code to crash when linked to Bartok's native sliding/compacting collector.

The fast verification times give us hope that there is still room to grow to support more features and better GC algorithms.  In particular, multithreading and pinning are essential for many applications and libraries.  Pinning should be easy for the mark-sweep collector, but would complicate the copying collector: pinned objects fragment the heap, forcing the allocator to allocate from a non-contiguous free space.
Multithreading would require reasoning about mutual exclusion (e.g. to keep allocators in different threads from allocating the same memory simultaneously), reasoning about suspending mutator threads during collection, and support for a more elaborate object pre-header word (for monitor operations on objects).
As the collectors grow, modularity becomes more important, so we're interested to see if the Boogie/Z3 approach can be combined with modular verification approaches based on separation logic and/or higher-order logic; hopefully, the automation provided by Boogie/Z3 will allow verification at a scale where modularity becomes essential.

\subsubsection*{Acknowledgments} The authors would like to thank Nikolaj Bj{\o}rner, Shaz Qadeer, Shuvendu Lahiri, Bjarne Steensgaard, Jeremy Condit, Juan Chen, Zhaozhong Ni, and the anonymous reviewers for many helpful discussions, suggestions, and comments.

\newcommand{\etalchar}[1]{$^{#1}$}

\end{document}